\documentclass[%
reprint,
superscriptaddress,
amsmath,amssymb,
prb,
]{revtex4-2}

\usepackage{graphicx}% Include figure files
\usepackage{dcolumn}% Align table columns on decimal point
\usepackage{bm}% bold math
\usepackage{multirow}
\usepackage{color}
\usepackage{ulem}

\begin{document}
	
	\preprint{APS/123-QED}
	
	\title{Spin-Orbit Coupled Insulators and Metals on the Verge of Kitaev Spin Liquids in Ilmenite Heterostructures}% Force line breaks with \\
	%\thanks{A footnote to the article title}%
	
	\author{Yi-Feng Zhao}
	\email{zyf@g.ecc.u-tokyo.ac.jp}
	\affiliation{Department of Applied Physics, University of Tokyo, Bunkyo, Tokyo 113-8656, Japan}
	\author{Seong-Hoon Jang}
	\affiliation{Institute for Materials Research, Tohoku University, Aoba, Sendai, 980-8577, Japan}

	\author{Yukitoshi Motome}
	\email{motome@ap.t.u-tokyo.ac.jp}
	\affiliation{Department of Applied Physics, University of Tokyo, Bunkyo, Tokyo 113-8656, Japan}

	\begin{abstract}
Competition and cooperation between electron correlation and relativistic spin-orbit coupling give rise to diverse exotic quantum phenomena in solids. 
An illustrative example is spin-orbit entangled quantum liquids, which exhibit remarkable features such as topological orders and fractional excitations. 
The Kitaev honeycomb model realizes such interesting states, called the Kitaev spin liquids, but its experimental feasibility is still challenging. 
Here we theoretically investigate hexagonal heterostructures including a candidate for the Kitaev magnets, an ilmenite oxide MgIrO$_3$, to actively manipulate the electronic and magnetic properties toward the realization of the Kitaev spin liquids. 
For three different structure types of ilmenite bilayers MgIrO$_3$/$A$TiO$_3$ with $A=$ Mn, Fe, Co, and Ni, we obtain the optimized lattice structures, %\blue{
the electronic band structures%}
, the stable magnetic orders, and the effective magnetic couplings, by combining {\it ab initio} calculations and the effective model approaches. 
We find that the spin-orbital coupled bands characterized by the pseudospin $j_{\rm eff}=1/2$, crucially important for the Kitaev-type interactions, are retained in the MgIrO$_3$ layer for all the heterostructures, but the magnetic state and the band gap depend on the types of %\blueout{the} 
heterostructures as well as the $A$ atoms. 
In particular, one type becomes metallic irrespective of $A$, while the other two are mostly insulating. 
We show that the insulating cases provide spin-orbit coupled Mott insulating states with 
dominant Kitaev-type interactions, accompanied by
different combinations of subdominant interactions depending on the %\blue{
heterostructural %} 
type and $A$, while the metallic cases  %\blue{
realize spin-orbit coupled metals 
with various doping %}
rates.
Our results indicate that these hexagonal heterostructures are a good platform %\blue{
for %} 
engineering 
%\blue{
electronic and %} 
magnetic properties of the spin-orbital coupled correlated materials, including the possibility of Majorana Fermi surfaces and topological superconductivity.

	\end{abstract}
	
	%\keywords{Suggested keywords}%Use showkeys class option if keyword
	%display desired
	\maketitle
	
	%\tableofcontents
	
	\section{INTRODUCTION}
	Strong electron correlations, represented as Coulomb repulsion $U$, play a pivotal role in 3$d$ transition metal compounds and lead to a plethora of intriguing phenomena, such as the Mott 
	transition and
	high-temperature superconductivity~\cite{Mott1968,Imada1998}.
	The other key concept of quantum materials, the spin-orbit coupling (SOC), represented as $\lambda$, is a
	relativistic effect entangling
	the spin degree of freedom and the orbital motion of electrons, which is an essential ingredient in the topological insulators~\cite{Qi2011,Hasan2010}. 
	Beyond their independent effects, 
	synergy between $U$ and $\lambda$
	has attracted increasing attention recently due to the emergence of 
	new states of matter, such as axion insulators~\cite{Wan2012,nenno2020axion} 
	and topological semimetals~\cite{Wan2011,Burkov2011,Armitage2018}. 
	In general, it is difficult for the SOC to dramatically influence the electronic properties in 3$d$ transition metal compounds 
	since $\lambda$ is much smaller than $U$.
	However, 
	when proceeding to 4$d$ and 5$d$ systems,
	the $d$ orbitals are spatially more spread out, which reduces $U$, and at the same time, the relativistic effect becomes larger for heavier atoms, enhancing $\lambda$. 
	Hence, in these systems, the competition and cooperation between 
	$U$ and $\lambda$ play %\blueout{s} 
	a decisive role in their
	electronic states and allow us to access the intriguing regime that yields the exotic correlated states of matter~\cite{witczak2014correlated}.
	
	%The most 
	One of the striking examples is 
	the spin-orbit coupled Mott insulator, typically realized in the
	iridium oxides with Ir$^{4+}$ valence, e.g., Sr$_2$IrO$_4$~\cite{Kim2008,Moon2008}. 
	In each Ir ion located in the center of the IrO$_6$ octahedr%\blueout{on}\blue{
	on%}
	, the crystal field energy, which is significantly larger than $U$ and 
	$\lambda$, 
	splits the $d$ 
	orbital manifold into low-energy $t_{\rm{2g}}$ 
	and high-energy $e_{\rm g}$ 
	ones. For Sr$_2$IrO$_4$, five electrons occupy the $t_{\rm{2g}}$ orbitals and make the system yield the $t_{\rm{2g}}^5$ low-spin state. Usually, the 
	partially-filled orbital 
	causes a metallic state according to the 
	conventional band theory, but
	the insulating state was observed in experiments~\cite{Crawford1994}. Considering the large SOC, the $t_{\rm{2g}}$ 
	manifold continues to split into high-energy doublet characterized with the pesudospin $j_{\rm{eff}}
	= 1/2$ and low-energy quartet with $j_{\rm{eff}}
	= 3/2$; 
	the latter is fully occupied and the former is half filled. 
	Finally, a Mott gap is opened in the half-filled $j_{\rm{eff}}=1/2$ band by $U$. 
	This accounts for the insulating nature of the system, and the Mott insulating state realized in the spin-orbital coupled bands is called the spin-orbit coupled Mott insulator.
	
	The quantum
	spin liquid (QSL), one of the most exotic quantum 
	states
	in the spin-orbit coupled Mott insulators, has received increasing attention due to the emergence of remarkable properties, e.g., fractional excitations~%\blue{
	\cite{Read1989} %} 
	and topological orders~\cite{wen1991topological}. In the QSL, long-range magnetic ordering is suppressed down to zero temperature
	due to 
	strong quantum fluctuations, though
	the localized magnetic moments are 
	quantum entangled~%\blue{
	\cite{anderson1973resonating,balents2010spin,zhou2017,broholm2020quantum}. The presence of fractional quasiparticles that obey the nonabelian statistics is not only of great fundamental physical research, but also promising toward quantum computation~\cite{Nayak2008}. In general, one route to realizing the QSL depends on geometrical frustration. Indeed, 
	experiments have evidenced several candidates of QSL 
	in antiferromagnets with lattice structures including triangular unit, where magnetic frustration is the common feature~%\blue{
	\cite{anderson1956,sachdev1992,ramirez1994strongly}%}
	. 
	The other route to the QSL is 
	the so-called exchange 
	frustration
	caused by the conflicting constraints 
	between anisotrpic 
	exchange interactions~%\blue{
	\cite{nussinov2015}%}
	.
	The strong spin-orbital entanglement in the spin-orbit %\blue{
	coupled %} 
	Mott insulators, in general, gives rise to spin anisotropy, offering a good playground for the exchange frustration, even on the lattices without geometrical frustration.
	
	The Kitaev model is a distinctive quantum spin model realizing the exchange frustration~\cite{kitaev2006anyons}. The model is defined for 
	$S$ = $\frac{1}{2}$ local magnetic moments 
	on the two-dimensional (2D) honeycomb lattice with the Ising-type bond-dependent anisotropic interactions, whose 
	Hamiltonian is 
	given by 
	\begin{equation}
		\mathcal{H} = \sum_{\alpha}\sum_{{\langle i,j \rangle}_{\alpha}}K_{\alpha}S_i^{\alpha}S_j^{\alpha}, 
	\end{equation}
	where $\alpha = x, y, z$ denote the three bonds of the honeycomb lattice, and 
	$K_{\alpha}$ is the 
	exchange coupling constant on the $\alpha$ bond; 
	the sum of ${\langle i,j \rangle}_{\alpha}$ 
	is taken for
	nearest-neighbor sites of $i$ and $j$ on the $\alpha$ bonds.
	In this model, the orthogonal anisotropies on the $x, y, z$ bonds provide the
	exchange frustration. 
	Importantly, the model is exactly solvable, and the ground state is a QSL 
	with fractional excitations,
	itinerant Majorana fermions and localized Z$_2$ fluxes~\cite{kitaev2006anyons}. Moreover, the anyonic excitations in %such physical systems 
	this model hold promise for applications in quantum computing~\cite{kitaev2003fault,kitaev2006anyons}; %. \blue{The Abelian anyons typically emerge in the gapped phase. In contrast, 
	especially, non-Abelian anyons, which follow braiding rules similar to those of conformal blocks for the Ising model, %occur in the gapless phase %but can be induced a gap 
	appear under an external %the influence of the 
	magnetic field%}
	.

	The exchange frustration in the Kitaev-type interactions can be realized in real materials when two conditions are met~\cite{Jackeli2009}. First,
	at each magnetic ion, the spin-orbit coupled Mott insulating state with pseudospin $j_{\rm{eff}}=1/2$ should be realized.
	Second, the pseudospins need to interact with each other through the ligands shared by neighboring octahedra which forms edge-sharing network.
	Over the past decade, enormous efforts have been devoted to exploring the candidate materials for the Kitaev QSL that meet these conditions~%\blue{
	\cite{takagi2019concept,motome2020hunting,motome2020materials,trebst2022kitaev}.
	Fortunately, dominant Kitaev-type interactions were indeed discovered in several materials,
	such as $A_2$IrO$_3$ with $A$ = Li
	and Na~%\blue{
	\cite{Singh2010,Singh2012,comin2012,sohn2013,hwan2015direct,das2019,zhao2021} %}
	and $\alpha$-RuCl$_3$~%\blue{
	\cite{Plumb2014,Kubota2015,sandilands2015,koitzsch2016,sinn2016electronic,do2017majorana,kasahara2018majorana,suzuki2021proximate}%}
	.
	Not only the 5$d$ and 4$d$ candidates, 3$d$ transition metal compounds like Co oxides have also been investigated~\cite{sano2018,Liu2018}. 
	In addition to these examples with ferromagnetic (FM) Kitaev interactions, the antiferromagnetic (AFM) Kitaev QSL candidates were also predicted by {\it ab initio} calculations, 
	e.g., 
	$f$-electron based magnets~\cite{Jang2019,Jang2020} and
	polar spin-orbit coupled Mott insulators $\alpha$-RuH$_{3/2}X_{3/2}$ with $X$ = Cl and Br in the Janus structure~\cite{Sugita2020}.
	
	Although
	a plethora of Kitaev QSL candidates have been 
	investigated,
	those 
	realizing the %\blue{ %} 
	Kitaev %\blueout{QSL}\blue{
	QSL in the ground state are still 
	missing, since a long-range magnetic order due to 
	parasitic interactions such as the Heisenberg 
	interaction 
	hinders the Kitaev QSL. 
	Considerable efforts have been dedicated to
	suppressing the parasitic interactions and/or
	enhancing the Kitaev-type interaction. One way is 
	to utilize heterostructures that incorporate the Kitaev candidates.
	For example, the 
	Kitaev interaction is promoted more than 50\% for 
	the heterostructure composed of %\blueout{two-dimensional (}
	2D %\blueout{)} 
	monolayers of $\alpha$-RuCl$_3$ and graphene compared to the pristine $\alpha$-RuCl$_3$, predicted by {\it ab initio} calculations~\cite{Biswas2019}. The heterostructures between a 2D $\alpha$-RuCl$_3$ and three-dimensional (3D) topological insulator BiSbTe$_{1.5}$Se$_{1.5}$ %\blue{ %} 
	evidenced the charge transfer phenomena, albeit the magnetic properties were not reported~\cite{Mandal2023}. Within the realm of Kitaev heterostructures, remarkably few studies have been designed for the composite 3D/3D superlattices due to the fabrication challenge~\cite{miura2020stabilization}.
	%. 
	To date, few attempts~%\blue{
	\cite{Kang2023} %} 
	have been made to investigate the development of the electronic band structure and the magnetic properties, particularly whether the %$K$ 
	Kitaev interaction is still dominant when constructing the 3D/3D heterostructures using Kitaev QSL candidates.

	In this paper, we theoretically study the electronic and magnetic properties in 
	bilayer heterostructures
	as an interface of 3D/3D superlattices
	using a recently-synthesized iridium ilmenite MgIrO$_3$~\cite{Haraguchi.2018} and other
	ilmenite magnets $A$TiO$_3$
	with $A$ = Mn, Fe, Co, and Ni as the substrate. This material 
	choice is motivated by two key considerations: (i) 
	All of these materials have been successfully synthesized in
	experiments, which is helpful for the  epitaxial growth of multilayer heterostructures, and (ii) the ilmenite MgIrO$_3$ 
	is identified as 
	a good candidate for Kitaev magnets~\cite{Jang2021,hao2022electronic}.~We consider
	three configurations of heterostructures, classified by type-I, II, and III, which are %valid 
	all chemically allowed due to the characteristics of the alternative layer stacking in ilmenites, as shown in Fig.~\ref{fig1}. The %\blue{
	electronic band structures%}
	, magnetic ground states, and the effective exchange interaction%\blue{
	s %} 
	are 
	systematically
	investigated by employing the combinatorial of 
	$ab~initio$ calculations, construction of
	the effective tight-binding model, and perturbation expansions. We find that (i) the spin-orbit coupled bands characterized by 
	the effective 
	pseudospin $j_{\rm{eff}}%
	= 1/2$, a key demand for Kitaev-type interactions, are still %remained 
	preserved in %\blue{
	the %} 
	MgIrO$_3$ layer %\blueout{s} 
	for all types of the heterostructures, (ii) type-I and III heterostructures  
	realize spin-orbit coupled Mott insulators excluding Mn type-III, 
	whereas type-II 
	ones are spin-orbit coupled metals with doped $j_{\rm{eff}}=1/2$ bands with various carrier concentrations, and (iii) 
	in almost all of the insulating cases, the Kitaev-type interactions are predominant, whereas the forms and magnitudes of the other parasitic interactions depend on the specific types of the heterostructures and the 
	$A$ atoms.

	The structure of the remaining article is as follows. In 
	Sec.~\ref{sec:IIa}, %\ref{structure}}
	we provide a detailed description of the optimized lattice structures of MgIrO$_3$/$A$TiO$_3$ heterostructures with $A$ = Mn, Fe, Co, and Ni. In %the rest of section
	Sec.~\ref{sec:IIb}, 
	%Sec.~\ref{sec:II}, 
	we introduce the methods employed in this work, including %\blue{
	the means for structural optimization and %} 
	%\red{[add the description of Sec.~\ref{structure}]}
	 the $ab~initio$ calculations with LDA+SOC+$U$ scheme~(Sec.~\ref{subsec:IIB}), %\blue{
	the estimates of the effective transfer integral and construction of the multiorbital Hubbard model (Sec.~\ref{subsec:IIC}), and the second-order perturbation that is used in the estimation of exchange interactions (Sec.~\ref{subsec:IID}). 
	In Sec.~\ref{sec:III}, we systematically display the results of the electronic band structures for three types of heterostructure MgIrO$_3$/$A$TiO$_3$. In Sec.~\ref{subsec:IIIA}, we present the electronic band structures for the paramagnetic state obtained by LDA+SOC, together with the projected density of states (PDOS) derived by the maximally-localized Wannier function (MLWF). In Sec.~\ref{subsec:IIIB}, we discuss the stable magnetic states within LDA+SOC+$U$ and show their band structures and PDOS. In Sec.~\ref{sec:IV}, we derive the effective exchange coupling constants for the heterostructures for which the LDA+SOC+$U$ calculations suggest spin-orbit coupled Mott insulating nature, and show their location in the phase diagram for the $K$-$J$-$\Gamma$ model. In %section
	Sec.~\ref{sec:VI}, we discuss the 
	%\blue{
	possibility of the realization of Majorana Fermi surfaces~(Sec.~\ref{subsec:VIA}) and exotic superconducting phases~(Sec.~\ref{subsec:VIB}) in %} 
	the heterostructures, and the feasibility of these heterostructures in  experiments~(Sec.~\ref{subsec:VIC}). 
	%\blue{
	Section~\ref{Sec:VII} is devoted to the summary %\blue{
	and prospects%}
	. %} 
	In Appendix~\ref{appendix:A}, %\blue{
	we present the details of the energy difference of the magnetic orders %}
	and the effective exchange couplings between the $A$ ions. We present additional information on %\blue{
	orbital projected band structures %} 
	for different %\blue{
	specific types of heterostructures %} 
	of $A$ atoms %\blue{
	in Appendix~\ref{appendix:B} and the band structures %\blueout{for}\blue{
	of %} 
	monolayer %\blueout{of} 
	MgIrO$_3$ in Appendix~\ref{appendix:C}%}
	. 
	%\section{\label{sec:II}CRYSTAL STRUCTURE AND METHODS}
	\section{\label{sec:IIa}HETEROSTRUCTURES}
	%\red{[throughout the method section, do we unify the tenses with the past tense? or present tense?]} 
	%\subsection{\label{structure}Crystal sturcture}
	
		\begin{figure}
		\includegraphics[width=1.0\columnwidth]{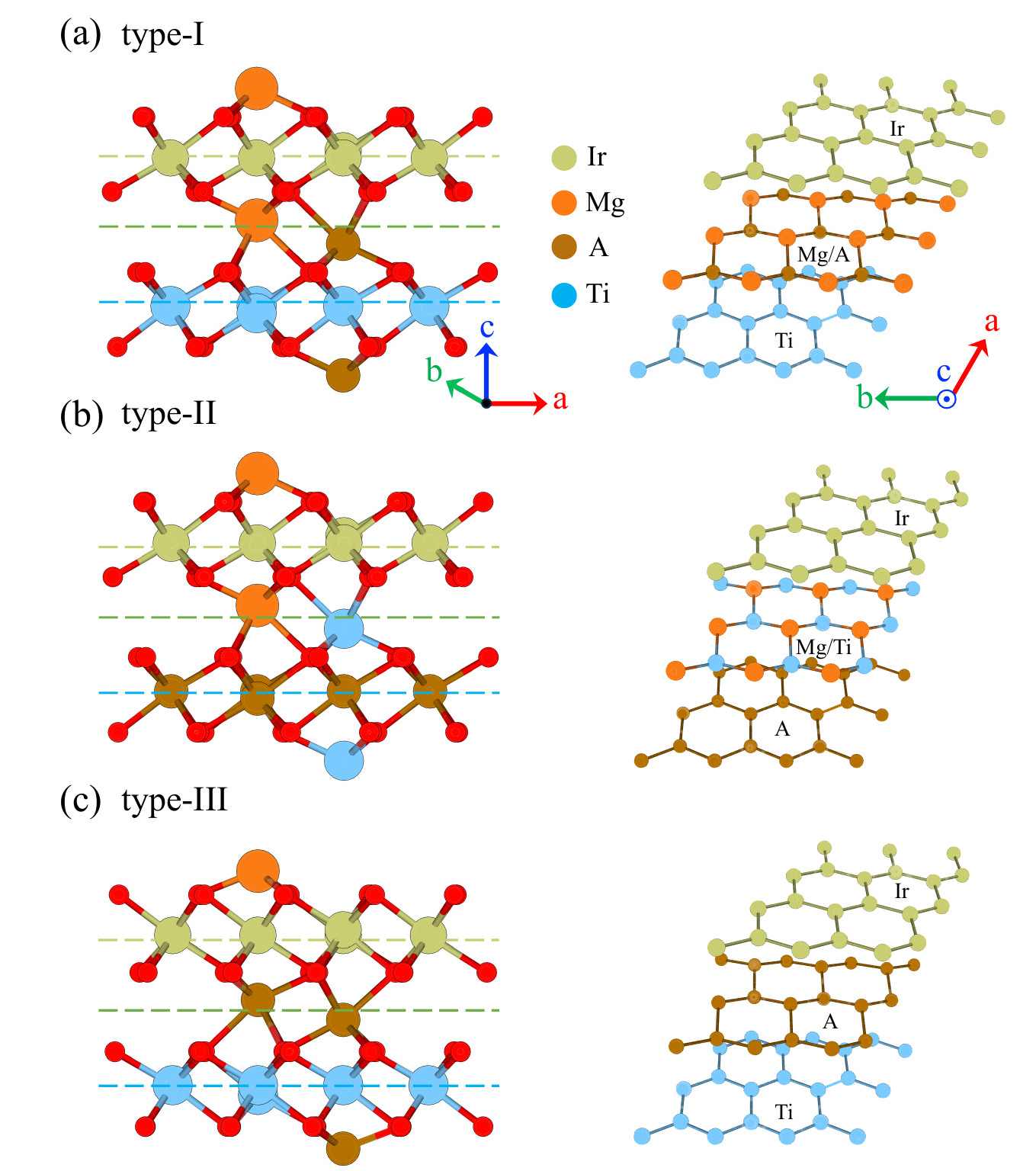}
		\caption{\label{fig1}%Side views (left) and bird's-eye views (right) 
			Schematic pictures of crystal structures for three types of the heterostructures MgIrO$_3$/$A$TiO$_3$: (a) type-I, (b) type-II, and (c) type-III with $A$ = Mn, Fe, Co, or Ni. The left and right panels show the side views and the bird's-eye views, respectively. The type-I is composed of the top honeycomb layer with edge-sharing IrO$_6$ octahedra and the bottom honeycomb layer of TiO$_6$,  sandwiching a honeycomb layer of alternating MgO$_6$ and $A$O$_6$ octahedra. 
			In the type-II, the bottom is replaced by the $A$O$_6$ honeycomb layer, leaving a mixture of MgO$_6$ and TiO$_6$ in the middle, %; 
			and in the type-III, the middle is replaced by the $A$O$_6$ honeycomb layer. 
			In type-I and II, the chemical formula is commonly given by %\blue{ 
			Mg$_2$Ir$_2$O$_6$/$A_2$Ti$_2$O$_6$, but that for type-III is Mg$A$Ir$_2$O$_6$/$A_2$Ti$_2$O$_6$. The crystal structures are embodied by VESTA \cite{momma2011vesta}.
			%\red{[If you used VESTA, we need to refer to it.]} 
			}
	\end{figure}
	
	%The structure of both 
	MgIrO$_3$ and $A$TiO$_3$
	both belong to ilmenite oxides $AB$O$_3$ with $R$$\bar{3}$ space group. 
	The lattice structure consists of alternative stacking of honeycomb layers with edge-sharing $A$O$_6$ octahedra and those with $B$O$_6$ octahedra. The common stacking layer structures reduce the lattice mismatch to form the heterostructures and also make them  feasible to fabricate in experiments. In this study, we consider heterostructures composed of monolayer of MgIrO$_3$ and $A$TiO$_3$ with the balance chemical formula, %is on purpose to constructed since we primarily focus on 
	to clarify the interface effect on the electronic properties of 3D/3D superlattices. 
	Specifically, we construct three types of heterostructures, distinguished by the intersurface atoms and pertinent octahedra in the middle layer, labeled as type-I, II, and III and shown in Fig.~\ref{fig1}.  For type-I, the top and bottom layers are made of honeycomb networks of IrO$_6$ and TiO$_6$ octahedra, respectively, whereas the sandwiching honeycomb layer is formed of alternating MgO$_6$ and $A$O$_6$ octahedra. In the type-II, the bottom layer is replaced by $A$O$_6$ honeycomb layer, resulting in a mixture of MgO$_6$ and TiO$_6$ in the middle layer. The type-III has a similar constitution  of top and bottom layers to type-I, while the %intermediate mixture octahedra alternated 
	middle layer is fully composed of %by 
	$A$O$_6$ octahedra. %\blue{
	We intentionally design %e %\blueout{d} 
	these structures to balance their chemical valences and prevent the presence of redundant charges. This can be derived from the chemical formula for each type of the heterostructure, such as Mg$_2$Ir$_2$O$_6$/$A_2$Ti$_2$O$_6$ for type-I and II, and Mg$A$Ir$_2$O$_6$/$A_2$Ti$_2$O$_6$ for type-III, respectively. %}
	
	%\blue{
	%\blue{Through the optimization schema in Sec.~\ref{subsec:IIB}, we obtain stable heterostructures.}
	We optimize %\blueout{d} 
	the lattice structures of the heterostructures by the optimization scheme in Sec.~\ref{subsec:IIB}. 
	The %structural 
	information of the stable lattice structures, including the in-plane lattice constant, the bond distance between adjacent Ir atoms and O atoms, and the angle between the neighboring Ir, O, and Ir atoms, are listed in Table~\ref{tab:table1} for three types of the heterostructures with different $A$ atoms. %\blue{
	For comparison, the experimental structures of the bulk MgIrO$_3$ are also listed%}
	. We find all the in-plane constants are close to the bulk value of 5.158~\AA, in which the maximum and minimum lattice mismatch is 1.2\% and 0.1\% respectively of type-II for the Fe atom and type-III for the Mn atom. See also the discussion in Sec.~\ref{subsec:VIC}. Meanwhile, not only the in-plane constants but also the bond distances of Ir atoms are both enlarged as the increase of ionic radii of $A$ atoms. The heterostructural %e 
	type
	can significantly influence the bond distance and angle as well. For example, the angle between neighboring Ir atoms and the 
	intermediate O atom $\theta$$_{\rm {Ir-O-Ir}} %$ 
	= 96.69%$
	^\circ$ of type-II for the Ni case largely increases from that of 94.03$^\circ$ of the bulk case. In terms of the Ir-Ir bond length ($d_{\rm {Ir-Ir}}$), the length of 2.986~\AA~for the bulk~\cite{Haraguchi.2018} 
	is substantially decreased to 2.930 \AA~of type-II for the Fe case. %But for 
	Meanwhile, the Ir-O bond length $d_{\rm {Ir-O}}$ of all 
	cases are enlarged compared with that of the bulk system of 1.942~\AA, in which type-III %for 
	with Co atoms is maximally influenced.
	
		\begin{table}[!h]
		\caption{\label{tab:table1}
			Structural information of optimized heterostructures for 
			MgIrO$_3$/$A$TiO$_3$ ($A$ = Mn, Fe, Co, and Ni): $a$ denotes the in-plane lattice constant, and $d$ and $\theta$ represent the bond distance and the angle between neighboring ions, respectively. %~\blue{
			The experimental information on the bulk %system %as a valuable reference 
			MgIrO$_3$ is also shown for comparison. %}a
		}
		\begin{ruledtabular}
			\begin{tabular}{cccccc%c
				}
				%&
				$A$&type&$a(\rm{\AA})$&$d_{\rm Ir-Ir} (\rm{\AA})$ &$d_{\rm Ir-O}(\rm{\AA})$ & $\theta_{\rm Ir-O-Ir}(^\circ)$ \\
				\hline
				%&
				\multirow{3}{*}{Mn}&I&5.167&2.986&1.997&95.92\\
				%~&
				~&II&5.127&2.962&2.009&95.57\\
				%~ &
				~&III&5.152&2.977&1.985&95.96\\
				\hline
				%&
				\multirow{3}{*}{Fe}&I&5.104&2.951&2.012&94.32\\
				%~&
				~&II&5.068&2.930&2.004&94.00\\
				%~&
				~&III&5.083&2.940&2.008&93.98\\
				\hline			
				%&
				\multirow{3}{*}{Co}&I&5.116&2.961&1.999&94.76\\
				%~&
				~&II&5.113&2.955&1.990&95.26\\
				%~&
				~&III&5.115&2.960&2.019&94.31\\			
				\hline
				%&
				\multirow{3}{*}{Ni}&I&5.148&2.977&2.006&95.03\\
				%~&
				~&II&5.181&2.994&1.997&96.69\\
				%~&
				~&III&5.173&2.994&1.995&94.18\\
				\hline
				%&%\blue{
				bulk~\cite{Haraguchi.2018}%}
				& %\blue{
			 %} 
				&%\blue{
				5.158%}
				&%\blue{
				2.986%}
				&%\blue{
				1.942%}
				&%\blue{
				94.03%} 
				\\
				
			\end{tabular}			
		\end{ruledtabular}
	\end{table}

	\section{\label{sec:IIb}METHODS}

	\subsection{\label{subsec:IIB}$\bm{%a
	Ab~initio}$ calculations}
	In the $ab~initio$ calculations, we use %\blueout{d} 
	the QUANTUM ESPRESSO~\cite{giannozzi2009quantum} based on the density functional theory~\cite{DFT1964}.  %\blueout{(DFT)%}~\cite{DFT1964}. 
	The exchange-correlation potential %\blue{
	is %} 
	treated as Perdew-Zunger functional by using the projector-augmented-wave method~\cite{PZ1981,BlochlPAW1994}. Under the consideration of the SOC effect, the fully relativistic functional %\blue{
	is %} 
	utilized for all the atoms except oxygens~\cite{Corso2014}. 
	%\blue{
	To obtain stable structures for heterostructures, we initially construct a bilayer MgIrO$_3$ structure using the experimental structure for the bulk material. Subsequently, we replace the lower half with $A$TiO$_3$ layer to create three different types of heterostructures in Fig.~\ref{fig1}. Then, we perform full optimization for both lattice parameters and the position of each ion until the residual force becomes less than 0.0001~Ry/Bohr. During the optimization procedure, the structural symmetry is retained as $R\bar{3}$ space group. %} 
	The 
	20~\AA~thick~vacuum %\blue{
	is %} 
	adopted to eliminate the interaction between adjacent layers. The 6$\times$6$\times$1 and 12$\times$12$\times$1 Monkhorst-Pack $k$-points meshes %\blue{
	are %}
	utilized for %geometry \blue{
	the structural optimization %} 
	and self-consistent calculations, respectively~\cite{Monkhorst1976}. The self-consistent convergence %\blue{
	is %} 
	set to 10$^{-8}$~Ry and the kinetic energy %\blue{
	is %} 
	chosen to 80~Ry for all the structural configurations, which are %both 
	respectively small and large enough to guarantee accurate results. To simulate the electron correlation effects for 3$d$ electrons of $A$ atoms and 5$d$ electrons of Ir atom, we adopt the %in 
	LDA+SOC+$U$ calculations~\cite{Liechtenstein1995} %, we parameterized the 
	with the Coulomb repulsions $U_A %$ 
	= 5.0$~eV, 5.3~eV, 4.5~eV, and 6.45~eV with $A$ = Mn, Fe, Co, and Ni atoms, respectively, and $U_{\rm {Ir}}%$ 
	= 3.0$~eV, accompanying with the Hund's-rule coupling with $J_{H}$/$U$ = 0.1 according to previous works~\cite{Chittari2016,Arruabarrena2022}.

	Based on the {\it ab initio} results, we also 
	obtain %\blueout{ed} 
	the MLWFs by using %, where 
	the $k$~points increased to 18$\times$18$\times$1 within the Momkhorst-Pack scheme~\cite{Monkhorst1976}. We select %\blueout{ed} 
	the $t_{2g}$, $2p$, and $3d$ orbitals respectively of Ir, O, and $A$ atoms to construct the MLWFs by employing the Wannier90~\cite{mostofi2014updated}. Herein, we include %\blueout{d} 
	O $2p$ and $A$ %\blueout{5}\blue{3}
	3$d$ orbitals due to their significant contribution near the Fermi level, as detailed in Figs.~\ref{fig2}-\ref{fig5}. %\blue{
	By utilizing the %} 
	MLWFs, we %\blue{
	construct the tight-binding models and calculate their band structures for comparison. %} 
	We also calculate the PDOS of each atomic orbital, including the effective angular momentum of Ir atoms $j_{\rm{eff}}$, from the tight-binding models. We consider the non-relativistic {\it ab initio} calculations and relative MLWFs for the estimation of transfer integrals (see Sec.~\ref{subsec:IIC}).%}

	\subsection{\label{subsec:IIC}Multiorbital Hubbard model}
	
	To estimate the effective exchange interactions between the magnetic Ir ions, we need the effective transfer integrals between neighboring Ir $t_{2g}$ orbitals with the association of O $2p$ orbitals by constructing MLWFs with LDA calculation in the paramagnetic state. It is noticeable that the effects of relativistic SOC and electron correlation are not taken into account in this calculation to circumvent the doublecounting in constructing the %later 
	effective spin models. %Thus
	Specifically, the effective transfer integral $t$ is %written down 
	estimated as~%\blue{
		\cite{Jang2021} %}
	\begin{equation}
		t_{{iu},{jv}} + \sum_{p}\frac{t_{{iu},{p}}t_{{jv},{p}}^*}{\Delta_{p-uv}}.
		\label{eq1}
	\end{equation}
	The first term denotes the direct hopping between two adjacent Ir atoms, 
	where $t_{{iu},{jv}}$ represents the transfer integral between orbital $u$ at site $i$ and orbital $v$ at site $j$. The second term denotes the indirect hopping between the two Ir atoms via the shared O 2$p$ orbitals, %\blue{
		where $t_{{iu},p}$ represents the transfer integral between Ir atom $u$ orbital at site $i$ 
		and  ligand atom $p$ orbital, and 
	$\Delta_{p-uv}$ is the harmonic mean of the energy of $u$ and $v$ orbitals measured from that of $p$ orbitals. Herein, we consider only hopping processes between the nearest-neighbor Ir atoms.%}
		
	%For 
	Using the effective transfer integrals, we construct a multiorbital Hubbard model with one hole occupying the $t_{2g}$ orbitals, whose 
	Hamiltonian  
	is given by
	\begin{equation}
		H = H_{\rm {hop}} + H_{\rm {tri}} + H_{\rm {soc}} + H_{U}.
		\label{eq:H}
	\end{equation}
	The first term denotes the kinetic energy of the $t_{2g}$ electrons as
	\begin{equation}
		H_{\rm{hop}} = -\sum_{i,j}\bm{c}_{i}^{\dagger}(\hat{T}_{ij}^{\gamma}\otimes{\sigma}_0)\bm{c}_j,
	\end{equation}
	where the matrix $\hat{T}_{ij}^{\gamma}$  
	includes the effective transfer integrals estimated by Eq.~\eqref{eq1}, $\gamma$ is the $x$, $y$, and $z$ bond connected by neighboring sites $i$ and $j$ which belong to different honeycomb sublattices, and $\sigma_0$ denotes the 2$\times$2 identity matrix; 
	$\bm{c}_{i}^{\dagger}$ = ($c_{i,yz,\uparrow}^{\dagger}$, $c_{i,yz,\downarrow}^{\dagger}$, $c_{i,zx,\uparrow}^{\dagger}$, $c_{i,zx,\downarrow}^{\dagger}$, $c_{i,xy,\uparrow}^{\dagger}$, $c_{i,xy,\downarrow}^{\dagger}$) denote the creation of one hole %on $i$ site 
	in the $t_{\rm {2g}}$ orbitals ($yz$, $zx$, and $xy$) carrying spin up ($\uparrow$) or down ($\downarrow$) at site $i$. %\blue{
	The second term in Eq.~\eqref{eq:H} denotes %T
	the trigonal crystal splitting as %Hamiltonian is given by}
	\begin{equation}
		H_{\rm{tri}} = -\sum_{i}\bm{c}_{i}^{\dagger}(\hat{T}_{\rm{tri}}\otimes\sigma_0)\bm{c}_{i},
	\end{equation}
	with $\hat{T}_{\rm{tri}}$ in the form of
	\begin{equation}
		\hat{T}_{\rm{tri}} = 
		\begin{pmatrix}
			0 & \Delta_{\rm{tri}} & \Delta_{\rm{tri}} \\
			\Delta_{\rm{tri}} & 0 & \Delta_{\rm{tri}} \\
			\Delta_{\rm{tri}} &\Delta_{\rm{tri}} & 0 \\
		\end{pmatrix}.
	\end{equation}
	The third term %are denoted as 
	denotes the SOC as 
	\begin{equation}
		H_{\rm{soc}} = -\frac{\lambda}{2}\sum_{i}\bm{c}_{i}^{\dagger}
		\begin{pmatrix}
			0 & i\sigma_z&-i\sigma_y \\
			-i\sigma_z&0&i\sigma_x \\
			i\sigma_y&-i\sigma_x&0 \\
		\end{pmatrix}
		\bm{c}_{i},
	\end{equation}
	where $\sigma$$_{\lbrace x,y,z \rbrace}$ are Pauli matrices, and %. 
	$\lambda$ is the SOC coefficient; %of atoms, 
	for instance, %that 
	$\lambda$ of Ir atom is %$\lambda_{\rm {Ir}}$ = 
	estimated at about 0.4~eV~\cite{Yamaji2014,Kim2014}. 
	The last term denotes the onsite Coulomb interactions %term $H_{U}$ is written 
	as~\cite{Ole1983,Roth1966} 
	\begin{eqnarray}
		H_{U}=&&\sum_{i}{U}n_{iu\uparrow}n_{iu\downarrow} \nonumber\\
		& &+\sum_{i,u<v,\sigma}[U'n_{iu\sigma}n_{iv\bar{\sigma}}+(U'-J_{\rm H})n_{iu\sigma}n_{iv\sigma}] \nonumber\\
		& &+\sum_{i,u\neq v}J_{\rm H} (c_{iu\uparrow}^{\dagger}c_{iv\downarrow}^{\dagger}c_{iu\downarrow}c_{iv\uparrow}+c_{iu\uparrow}^{\dagger}c_{iu\downarrow}^{\dagger}c_{iv\downarrow}c_{iv\uparrow}),
		\label{eq:HU}
	\end{eqnarray}
	with $n_{iu\sigma}$ = $c_{iu\sigma}^{\dagger}c_{iu\sigma}$; $\bar{\sigma} = \downarrow$ ($\uparrow$) for $\sigma = \uparrow$ ($\downarrow$). In Eq.~\eqref{eq:HU}, %T
	the first, second, and third summations represent %term is 
	the intraorbital Coulomb interaction in the same orbital with opposite spins, %. The second terms are 
	the interorbital Coulomb interactions between orbital $u$ and orbital $v$, and %with spins and $J_{\rm H}$ is the Hund's-rule coupling. The third term denotes 
	the spin-flip and pair-hopping processes, respectively.
	
	\subsection{\label{subsec:IID}%\blue{
	Second-order perturbation%}
	}
	
	%\blue{
	For Ir$^{5+}$ %atoms
	ions, the $t_{2g}$ ma%in
	nifold splits into a doublet %$j_{\rm {eff}}$ = 1/2 state 
	and a quartet under the SOC, %$j_{\rm {eff}}$ = 3/2 states
	which are respectively characterized by the pseudospin $j_{\rm {eff}}=1/2$ and $3/2$. %The could be described by the Kramers doublet with one hole occupied under the synergy of SOC and trigonal crystal splitting in Eq.~\eqref{eq:H}. 
	In the ground state, the latter is fully occupied and the former is half filled, which is described by %T
	the Kramers doublet $\left |j_{\rm{eff}} = 1/2, + \right >$ and $\left |j_{\rm{eff}} = 1/2, - \right >$~%\blue{
\cite{khaliullin2005,Jackeli2009}%}
: %are given as
	\begin{gather}
		\left |j_{\rm{eff}} = 1/2, + \right > = \frac{1}{\sqrt{3}}\left(|d_{yz} \downarrow \rangle + i%\frac{1}{\sqrt{3}}
|d_{zx} \downarrow \rangle + %\frac{1}{\sqrt{3}}
|d_{xy}\uparrow\rangle\right), \\
		\left |j_{\rm{eff}} = 1/2, - \right > = \frac{1}{\sqrt{3}}\left(|d_{yz} \uparrow \rangle - i%\frac{1}{\sqrt{3}}
|d_{zx} \uparrow \rangle - %\frac{1}{\sqrt{3}}
|d_{xy}\downarrow\rangle\right).
	\end{gather} %}
	
	%\blue{
	When the system is in the spin-orbit coupled Mott insulating state with the low-spin $d^5$ configuration, the low-energy physics can be described by the pseudospin $j_{\rm eff}=1/2$ degree of freedom. In this case, 
	the effective exchange interactions between the pseudospins can be estimated by using the second-order perturbation theory in the atomic limit, where the three terms in Eq.~\eqref{eq:H}, $H_{\rm {tri}} + H_{\rm {soc}} + H_{U}$, are regarded as unperturbed Hamiltonian, and $H_{\rm{hop}}$ is treated as perturbation. The  energy correction for a neighboring pseudospin pair in the second-order perturbation is given by
		\begin{equation}
			E_{\sigma_{i}^{\prime},\sigma_{j}^{\prime};\sigma_{i},\sigma_j}^{(2)}=\sum_{n}\frac{\langle \sigma_i^{\prime}
			\sigma_j^{\prime}|H_{\rm {hop}}|n\rangle\langle n|H_{\rm{hop}}|\sigma_i
			\sigma_j\rangle}{E_0-E_n}, 
			\label{eq:energy}
		\end{equation}
		where $\sigma_{i}$ and $\sigma_i^{\prime}$ denote the pseudospin $+$ or $-$ %up ($\uparrow$) and down ($\downarrow$) 
		at site $i$, $|\sigma_i\sigma_j\rangle$ and $\langle \sigma_i^{\prime}\sigma_j^{\prime}|$ is the initial and final states during the perturbation process, respectively, %\red{[why limited to the combination of $dd$ and $uu$?]} 
		and %. 
		$|n\rangle$ is the intermediate state with 5$d^4$-5$d^6$ or 5$d^6$-5$d^4$ electron configuration; %. 
		$E_0$ is the ground state energy for the 5$d^5$-5$d^5$ electron configuration, and $E_n$ is the energy eigenvalue for the intermediate state $|n\rangle$. Here, $|n\rangle$ and $E_n$ are obtained %B
		by diagonalizing the unperturbed Hamiltonian $H_{\rm {tri}} + H_{\rm {soc}} + H_{U}$. 
	
    %\blue{
	The effective pseudospin Hamiltonian is written in the form of 
	\begin{equation}
		H %_{\rm{K}}$ 
		= %$
		\sum\limits_{\gamma = x,y,z} \sum\limits_{\left\langle i,j\right\rangle}\bm{S}_i^{\rm T} \bm{J}_{ij}^{\gamma} \bm{S}_j
		,
	\end{equation}
	where $i,j$ denote the neighboring sites, and $\gamma$ denotes the three types of Ir-Ir bonds on the MgIrO$_6$ honeycomb layer that are related by $C_3$ rotation. The coupling constant $\bm{J}_{ij}^{\gamma}$ is explicitly given, e.g., for the $z$ bond as 
	\begin{equation}
		\bm J_{ij}^{z} =
		\begin{bmatrix}
			J & \Gamma &\Gamma^\prime \\
			\Gamma & J & \Gamma^\prime \\
			\Gamma^\prime & \Gamma^\prime & K
		\end{bmatrix},
	\end{equation}
	where $J$, $K$, $\Gamma$, and $\Gamma^\prime$ represent the coupling constants for the isotropic Heisenberg interaction, the bond-dependent Ising-like Kitaev interaction, and two types of the symmetric off-diagonal interactions. %}
	%
	%\blue{
	Using the perturbation energy $E_{\sigma_{i}^{\prime},\sigma_{j}^{\prime},\sigma_{i},\sigma_j}^{(2)}$ obtained by Eq.~\eqref{eq:energy}, the coupling constants are calculated as
	%	\begin{gather}
	\begin{align}
			&J=2E_{+,-;-,+}^{(2)}, \\
			&K = 2\left(E_{+,+;+,+}^{(2)}-E_{+,-;+,-}^{(2)}\right), \\
			&\Gamma = 2{\rm{Im}}\left\{E_{-,-;+,+}^{(2)}\right\}, \\
			&\Gamma^\prime = 4{\rm{Re}}\left\{E_{+,+;+,-}^{(2)}\right\}.
	%	\end{gather}
	\end{align}
	%}

	\section{\label{sec:III}ELECTRONIC BAND STRUCTURE}
	
	\subsection{\label{subsec:IIIA}LDA+SOC results for paramagnetic states}
			\begin{figure*}
		\includegraphics[width=2.0\columnwidth]{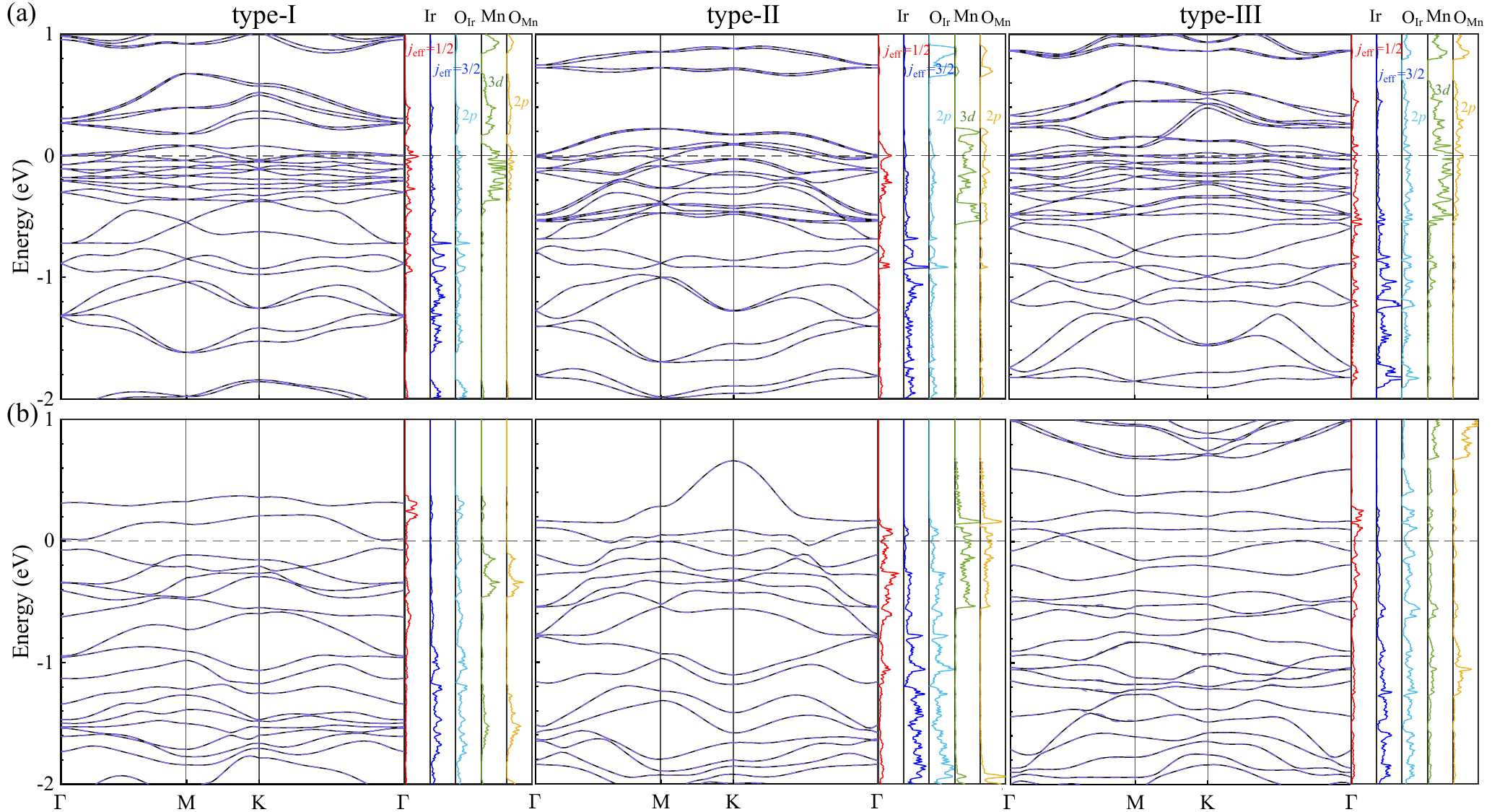}
		\caption{\label{fig2}The band structures of MgIrO$_3$/MnTiO$_3$ for type-I, II, and III with (a) the LDA+SOC calculations for the paramagnetic state and (b) the LDA+SOC+$U$ calculations for %\blueout{the N$\rm{\acute{e}}$el AFM state for both layers}\blue{
		the stable magnetic orders (see Sec.~\ref{SubsubIIIB:2})%}
		. The black lines represent the electronic structure obtained by %DFT 
		the {\it ab initio} calculations, and the light-blue dashed curves represent the electronic dispersions obtained by tight-binding parameters using the MLWFs. The right panels in each figure denote the %projected 
		PDOS for different orbitals on specific atoms: %\blue{
		The red and blue lines represent the $j_{\rm{eff}}$ manifolds of Ir atoms, the cyan and orange lines represent the $2p$ orbitals of O atoms in IrO$_6$ octahedra (O$_{\rm Ir}$) and MnO$_6$ octahedra (O$_{\rm Mn}$), respectively, and the green line represents the $3d$ orbitals of Mn atoms. %} 
		The Fermi energy is set to zero.}
	\end{figure*}
	
		\begin{figure*}
		\includegraphics[width=2.0\columnwidth]{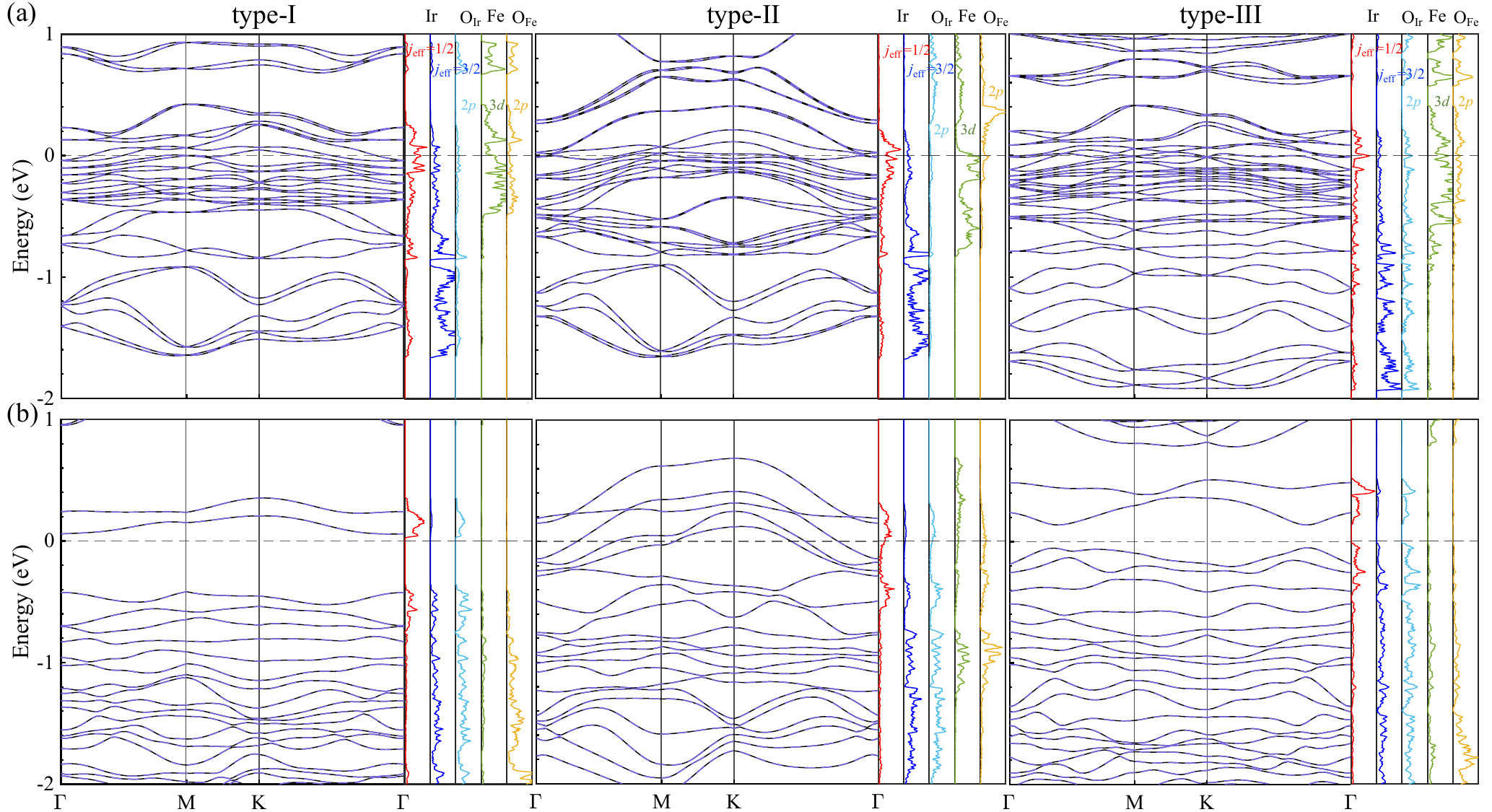}
		\caption{\label{fig3}The band structures of MgIrO$_3$/FeTiO$_3$ for type-I, II, and III obtained by (a) the LDA+SOC calculations for the paramagnetic state and (b) the LDA+SOC+$U$ calculations for %\blueout{the N$\rm{\acute{e}}$el AFM state for both layers}\blue{
		the stable magnetic orders (see Sec.~\ref{SubsubIIIB:2})%}
		. The notations are common to Fig.~\ref{fig2}.}
	\end{figure*}
	
	\begin{figure*}
		\includegraphics[width=2\columnwidth]{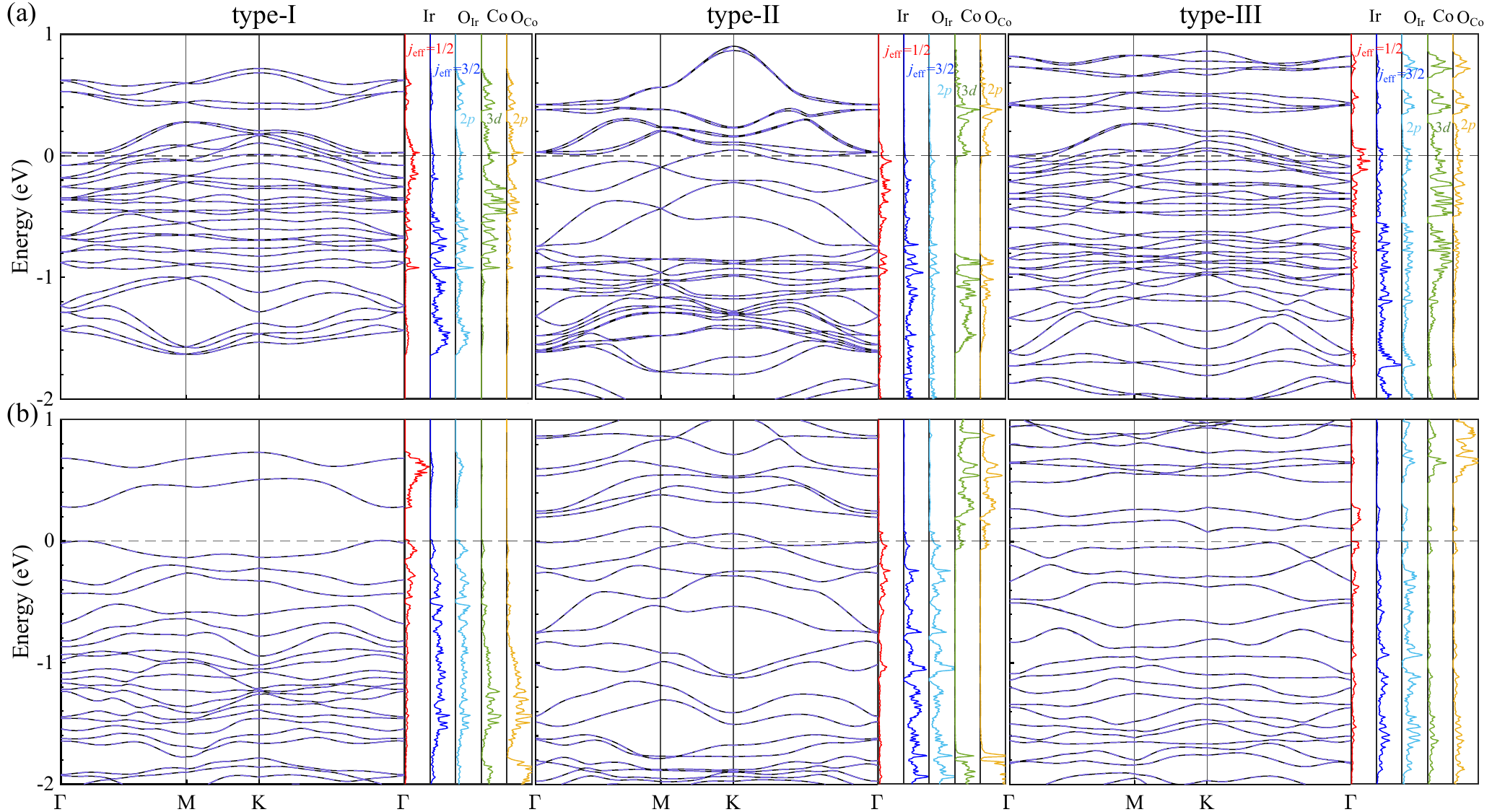}
		\caption{\label{fig4}The band structures of MgIrO$_3$/CoTiO$_3$ for type-I, II, and III obtained by (a) the LDA+SOC calculations for the paramagnetic state and (b) the LDA+SOC+$U$ calculations for %\blueout{the N$\rm{\acute{e}}$el AFM state for both layers}\blue{
		the stable magnetic orders (see Sec.~\ref{SubsubIIIB:2})%}
		. The notations are common to Fig.~\ref{fig2}.}
	\end{figure*}
	
	\begin{figure*}
		\includegraphics[width=2.0\columnwidth]{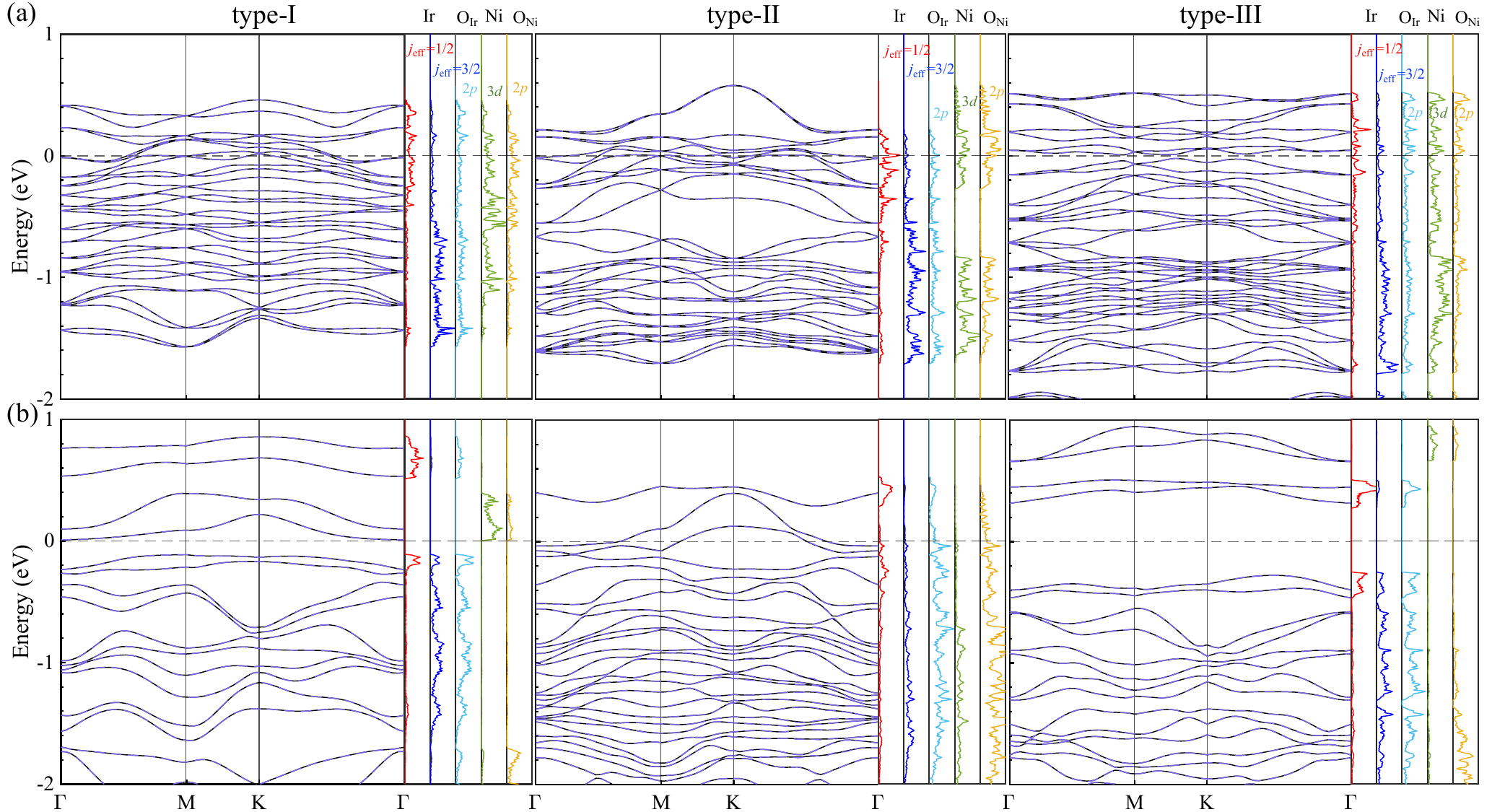}
		\caption{\label{fig5}The band structures of MgIrO$_3$/NiTiO$_3$ for type-I, II, and III obtained by (a) the LDA+SOC calculations for the paramagnetic state and (b) the LDA+SOC+$U$ calculations for %\blueout{the N$\rm{\acute{e}}$el AFM state for both layers}\blue{
		the stable magnetic orders (see Sec.~\ref{SubsubIIIB:2})%}
		. The notations are common to Fig.~\ref{fig2}.}
	\end{figure*}
	
	Let us begin with the electronic band structures from LDA+SOC calculations. %, as 
	The results for MgIrO$_3$/$A$TiO$_3$ with $A$ = Mn, Fe, Co, and Ni are shown in Figs.~\ref{fig2}(a), \ref{fig3}(a), \ref{fig4}(a), and %-
	\ref{fig5}(a), respectively. Here we display the band structures in the paramagnetic state obtained by the {\it ab initio} calculations (black solid lines) and the MLWF analysis (blue dashed lines), together with the atomic orbitals
	PDOS %\blue{
	including 5$d$ of Ir atoms, 3$d$ %\blue{
	of %} 
	$A$ atoms, and 2$p$ %\blue{
	of %} 
	O atoms related to IrO$_6$ and $A$O$_6$ octahedra. 
	It is obvious that the systems are metallic for all the types of heterostructures, regardless of the choice of $A$ atoms. 
	The strong SOC splits the $t_{\rm {2g}}$ bands of Ir atoms into $j_{\rm eff}=1/2$ and $j_{\rm{eff}}$ = 3/2 bands, as depicted in the PDOS. Specifically, the $j_{\rm eff}=1/2$ bands are predominated to form the metallic bands in the proximity of the Fermi level, covering the energy region almost from $-1.0$~eV to $0.2$~($0.4$)~eV for $A$ = Mn with type-II (type-I and III), from $-0.8$~eV to $0.2$~eV for $A$ = Fe with all types, from $-0.6$~($-1.0$)~eV to $0.2$~eV for $A$ = Co with type-II (type-I and III), and from $-0.4$~eV to $0.4$~eV for $A$ = Ni with type-I and II, and $-0.8$~eV to $0.2$~eV with type-III. %\blue{
	On the other hand, the $j_{\rm{eff}}$ = 3/2 bands primarily occupy the energy region below the  $j_{\rm eff}=1/2$ bands. 
	
	From the PDOS  %\blue{
	in right panels of Figs.~\ref{fig2}(a), \ref{fig3}(a), \ref{fig4}(a), and 
	\ref{fig5}(a)%}
	, the 3$d$ bands of $A$ atoms 
	and the O~$2p$ bands %\blue{
	of $A$O$_6$ octahedra simultaneously across the Fermi level%}
	, %\blue{
	with hybridization %} 
	with the Ir~$5d$ bands. %\red{[How do you judge the hybridization? Strictly speaking, the PDOS in the same energy range does not necessarily mean the hybridization; we need to look into the orbital components in each band.]}\blue{[I judge the hybridization of the bands from the overlap between the different orbital occupations. Although it can not quantitatively judge the hybridization, it is qualitative to state this. if you think these representations are not proper, we can change the state of `hybridization']} \red{[I believe in principle you can quantify the hybridization to check the orbital components of each band at each k point.]} 
	%\red{[Do we need to show the oxygen PDOS separately for the Ir and $A$ layers?]}\blue{[I have updated the PDOS with different octahedrals.]} \red{[I think the order Ir, O$_{\rm Ir}$, Mn, O$_A$ is better. You do not write ``Ir" twice (you can show it only once above between the two columns. Please add the description for O$_{\rm Ir}$ and O$_A$ in the caption.]}
	Notably, the energy range of the $A$ 
	3$d$ bands closely overlaps with that of the Ir 5$d$ %~\blue{
	$j_{\rm{eff}}=1/2$ bands for $A$ = Mn and Fe, but it overlaps with both $j_{\rm{eff}}=1/2$ and $j_{\rm{eff}}=3/2$ manifold for $A$ = Co and Ni. As to the O 2$p$ bands, the energy range of the PDOS overlaps with that for corresponding Ir $5d$ or $A~3d$ encapsulated in the octahedra, suggesting Ir-O and $A$-O hybridization. %}

	\subsection{\label{subsec:IIIB} LDA+SOC+$U$ results for magnetic states}
	\subsubsection{\label{subsubIIIB:1}Magnetic ground states}
	
	%\blue{
	The bulk %\blueout{states}\blue{
	counterpart %} 
	of each constituent of the heterostructures exhibit%\blue{
	s %} 
	some magnetic long-range orders in the ground state. In the bulk MgIrO$_3$, Ir ions show a zigzag-type AFM order with the magnetic moments lying almost within the honeycomb plane~\cite{hao2022electronic}. In the bulk $A$TiO$_3$, the $A$ = Mn and Fe ions show N\'eel-type AFM orders with out-of-plane magnetic moments~\cite{shirane1959neutron}, while the $A$ = Co and Ni ions support N\'eel orders with in-plane magnetic moments~\cite{newnham1964crystal,shirane1959neutron}. It is intriguing to examine how these magnetic orders in the bulk are affected by making heterostructures. 
	We determine the stable magnetic ground states for each heterostructure through {\it ab initio} calculations %}
	by including the effect of electron correlations based on the LDA+SOC+$U$ method.  To determine the potential magnetic ground state for each heterostructure, we compare the energy across a total of 16 magnetic configurations among all combinations of %\blueout{four} \blue{
	following %} 
	types of the magnetic orders: FM %\blue{
	and N\'eel %}
	orders with in-plane and out-of-plane %}
	magnetic moments for %\blue{
	$A$ layer%}
	, and %\blue{
    N\'eel and zigzag orders with in-plane magnetic moments, as well as FM with in-plane and out-of-plane magnetic moments for Ir layer, %}
	%N\'eel%\blue{
	%	\redout{,} and zigzag %} 
	%	with in-plane %\blueout{and out-of-plane }
	%	magnetic moments %\blue{
	%	for \red{the} Ir layer%}
	%	, %\blueout{and} \blue{
	%	as well as N\'eel and %}
	%	zigzag with in-plane magnetic moments %\blueout{,} 
	%	for %\blueout{Ir and }
	%	\red{the} $A$ layer%\blueout{s}\blue{
	within a 2$\times$2$\times$1 supercell setup. 
	%\red{[Does this correctly describe what you did? Did you try all four types for aboth Ir and $A$ layer?]} 
	%\red{[I am confused: Didn't you try out-of-plane N\'eel for the $A$ layer?]} \blue{[yes, I tried FM and N\'eel with in-plane and out-of-plane for $A$ layer, as well as FM with in-plane and out-of-plane and N\'eel and zigzag with in-plane for Ir layer.]}
	%\red{[For Ir layer, did you try both in-plane and out-of-plane FM, in addition to in-plane N\'eel and zigzag? (no specification of the moment direction for FM is found in the above description)]} \blue{[yes, I have tried them]}
	%\red{[I believe you tried FM and N\'eel with in and out moments for A, and N\'eel and zigzag with in moments for Ir. Is this ok? (your previous sentence excludes, for instance, N\'eel for A + N\'eel for Ir)]} 
	
	\begin{table}[htbp]
		\caption{\label{tab:table2}
			Stable magnetic orders obtained by the LDA+SOC+$U$ calculations: FM, N\'eel, and zigzag denotes the ferromagnetic, N\'eel-type antiferromagnetic, and zigzag-type antiferromagnetic orders, respectively. While the directions of the magnetic moments are all in-plane for the Ir layers, those for $A$ can be in-plane (``in") or out-of-plane (``out") depending on $A$ and type of the heterostructure.
		}
		\begin{ruledtabular}
			\begin{tabular}{ccccc%c
				}
				%&
				$A$&layer&type-I&II&III \\
				\hline
				%&
				\multirow{2}{*}{Mn}&Ir&in-zigzag&in-zigzag&in-zigzag\\
				%&~
				&Mn&in-N\'eel&out-FM&in-N\'eel\\
				\hline
				%&
				\multirow{2}{*}{Fe}&Ir&in-N\'eel&in-N\'eel&in-FM\\
				%&~
				&Fe&out-N\'eel&in-FM&in-FM\\
				\hline
				%&
				\multirow{2}{*}{Co}&Ir&out-FM&in-FM&in-N\'eel\\
				%&~
				&Co&in-N\'eel&in-N\'eel&in-N\'eel\\
				\hline
				%&
				\multirow{2}{*}{Ni}&Ir&in-N\'eel&in-zigzag&in-zigzag\\
				%&~
				&Ni&out-N\'eel&in-FM&in-N\'eel\\
			\end{tabular}			
		\end{ruledtabular}
	\end{table}
	
	The results of the most stable magnetic state are listed in Table~\ref{tab:table2}. The details of the energy comparison are shown in Appendix~\ref{appendix:A}. In most cases, the $A$ ions show N\'eel orders as in the bulk cases, but the direction of magnetic moments is changed from the bulk in some cases. For instance, type-I and III with $A$ = Mn and type-III with $A$ = Fe switch the moment direction from out-of-plane to in-plane. While all the types with $A$ = Co retain the in-plane N\'eel states, type-I with $A$ = Ni is changed into the out-of-plane N\'eel state. The results indicate that the direction of magnetic moments are sensitively altered by making the heterostructures with MgIrO$_3$. 
	%\red{[Can you comment on these magnetic orders from the viewpoint of the effective magnetic couplings between the $A$ ions?]} 
    Meanwhile, the other cases, type-II with $A$ = Mn, Fe, and Ni as well as type-III with $A$ = Fe, are stabilized in the FM state. %\blue{
	These results are in good %agreement 
	accordance with the effective magnetic couplings between the $A$ ions estimated by a similar perturbation theory in Sec.~\ref{subsec:IID} \cite{liechtenstein1987local,he2021tb2j}, attesting to the reliability of magnetic properties in heterostructures (see Appendix~\ref{appendix:A}). %} 
	%\redout{We will show in Sec.~\ref{SubsubIIIB:2} that all type-I and III, except for the type-III with $A$ = Mn, show insulating band structures with gap opening, whereas all type-II cases remain metallic as in the LDA+SOC results.} 
	
	The magnetic states in the Ir layer are more complex due to the possibility of the zigzag state. For $A$ = Mn, the magnetic ground states of the Ir layer in all three types prefer the in-plane zigzag state as in the bulk of MgIrO$_3$. In contrast, for $A$ = Fe,  type-I and II are stable in the in-plane N\'eel state, but type-III prefers the in-plane FM state. For $A$ = Co, only type-III stabilizes the in-plane N\'eel state, while others exhibit the out-of-plane FM state for type-I and the in-plane FM state for type-II. Lastly, for $A$ = Ni, both type-II and type-III prefer the in-plane zigzag state, while it changes into the in-plane %\blue{
	N\'eel state %} 
	in type-I. These results indicate that the magnetic state in the Ir honeycomb layer is susceptible to both $A$ ions and the heterostructure type. We will discuss this point from the viewpoint of the effective magnetic couplings in Sec.~\ref{sec:IV}.

	\subsubsection{\label{SubsubIIIB:2}Band structures}
	
	We present the band structures obtained by the LDA+SOC+$U$ calculations in Figs.~\ref{fig2}(b), \ref{fig3}(b), \ref{fig4}(b), and %-
	\ref{fig5}(b) for $A$ = Mn, Fe, Co, and Ni, respectively.
	In these calculations,
	we adopt the stable magnetic states in Table~\ref{tab:table2}, except for the cases with in-plane zigzag order in the Ir layer. For the zigzag cases, for simplicity, we replace them by the in-plane N\'eel solutions, keeping the $A$ layer the same as the stable one. This reduces significantly the computational cost of the MLWF analysis for the zigzag state with a larger supercell. We confirm that the band structures for the N\'eel state are similar to those for the zigzag state, and the energy differences between the two states are not large as shown in Appendix~\ref{appendix:A}. %} 
	%\red{[Is this shown in Appendix~\ref{appendix:A}?]} \blue{[yes, in Applendix~A, we show energy difference between all types of magnetic states. But without band structures with zigzag state.]}
	%\red{[Did you fix the magnetic moments? Or optimize their magnitudes and directions?]} \blue{[I just fix the magentic directions and optimize the magnitudes.]} \red{[How about the $A$O$_6$ layer?]}\blue{[It is also the N\'eel AFM state.]} 
	%\red{[to which direction did you fix the moments?]} 
	
\begin{figure}
	\includegraphics[width=1.0\columnwidth]{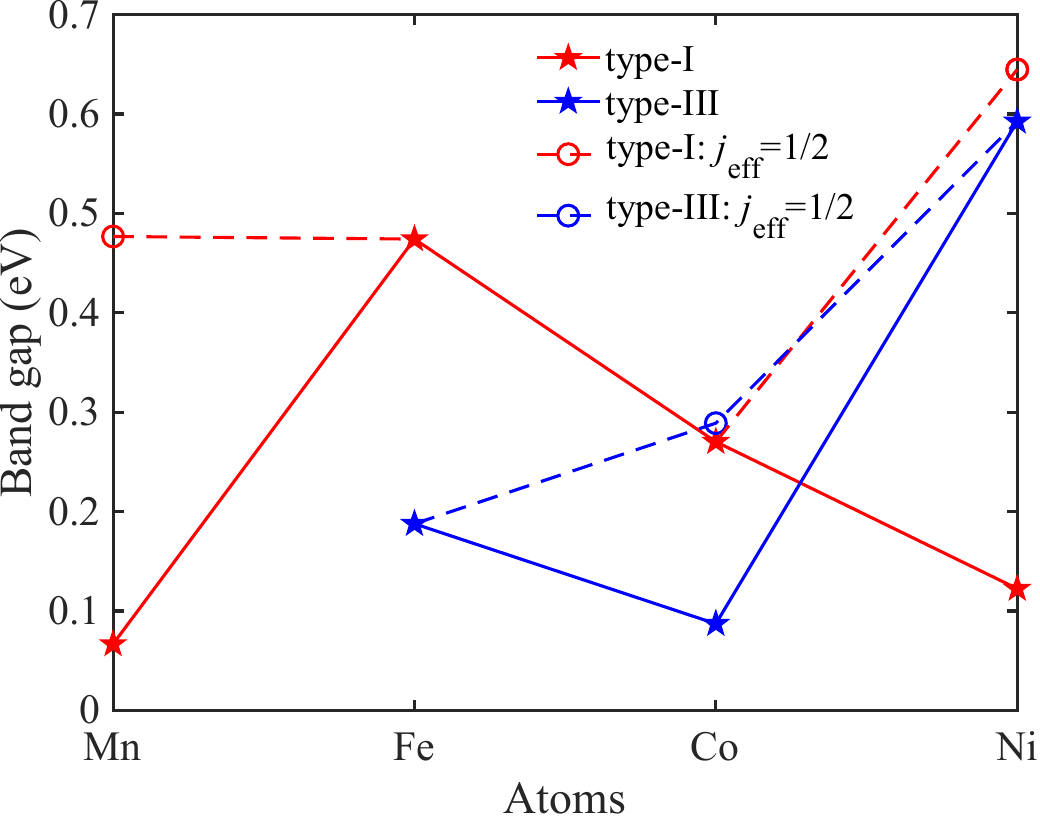}
	%\blue{
	\caption{\label{fig6}The %\blue{
	 %} 
	band gap in the insulating states for type-I and III obtained by the LDA+SOC+$U$ calculations. %\blue{
	The open circles represent the gaps opening in the $j_{\rm{eff}}=1/2$ bands for Mn and Ni of type-I and Co of type-III. %}
	%\red{[may we change the asterisks to, e.g., open circles for more clarity?]} 
	Note that the Co type-III is exceptional since the $j_{\rm{eff}}=1/2$ bands are at $3/4$ filling, rather than half filling in the other cases; see the text for details. 
	%\red{[open stars are difficult to distinguish from the closed ones]}
	%\red{[Can we plot the asterisk also for Co type-III (from Fig.~9(b))?]}	 
	%\red{[Did you confirm that the gap opens between the $j_{\rm{eff}}=1/2$ bands for Fe type-III (unlike Co type-III)?]} \blue{[Yes, I have checked it, the band gap is opened between $j_{\rm{eff}}=1/2$ bands, not the case like Co type-III.]}
	%\red{[We may connect the asterisks to the neighboring stars (e.g., Mn asterisk with Fe star) by dashed lines.]} 
	}
	\end{figure}

	%\red{[In the following, I think it is better to reorganize the paragraphs to discuss first the formation of the spin-orbit coupled Mott insulators in type I and III including the gap estimates, and then the appearance of the spin-orbit coupled metals in type II.]} 
	%\blue{
	%\blue{
	When we turn on Coulomb repulsions for both Ir and $A$ atoms, most of the type-I and III heterostructures become insulating, except for Mn type-III. The band gaps, %} 
	obtained by $E_{\rm g} =E_{\rm c}- E_{\rm v}$, are shown in Fig.~\ref{fig6}, where $E_{\rm c}$ denotes the energy of conduction band minimum and $E_{\rm v}$ is that of valence band maximum. %\blue{
	In all cases, except for type-I with $A$ = Mn and Ni and type-III with $A$ = Co, the gap is defined by the $j_{\rm{eff}}=1/2$ bands of Ir ions, that is, both conduction and valence bands are $j_{\rm{eff}}=1/2$, and the $j_{\rm{eff}}=1/2$ bands is half filled. It is worth highlighting that there are four $j_{\rm eff}=1/2$ bands, which originate from different sites of Ir atoms with opposite magnetic moments; in the bulk and monolayer cases they are degenerate in pair~\cite{Jang2021}, but the degeneracy is lifted in the heterostructures and two out of four are occupied in the half-filled insulating state. Meanwhile, in the cases of type-I with $A$ = Mn and Ni and type-III with $A$ = Co, the $3d$ bands of $A$ ions hybridized with O $2p$ orbitals intervene near the Fermi level, and the gap is defined between the $j_{\rm eff}=1/2$ and $3d$ bands. In these cases, however, a larger gap is well preserved in the $j_{\rm{eff}}=1/2$ bands, as shown in Figs.~\ref{fig2}(b), \ref{fig4}(b), and \ref{fig5}(b). %}
	We note that the Co type-III is a further exception since the gap opens between the highest-energy $j_{\rm{eff}}=1/2$ band and the Co $3d$ band; the $j_{\rm{eff}}=1/2$ bands are not half filled but $3/4$ filled (see Appendix~\ref{appendix:B}). 
	%\red{[Is this also the case for Co type-III? (Fig.~9(b))]} \blue{[Yes, it is also suit for Co type-III, but where the band gap is opened by two upper $j_{\rm{eff}}=1/2$ bands, not by the upper and lower $j_{\rm{eff}}=1/2$ bands. But anyway, I will add the band gap of $j_{\rm{eff}}=1/2$ bands in Fig. (6).]} 
	We plot %} 
	the band gap defined by the $j_{\rm{eff}}=1/2$ bands by red asterisks in Fig.~\ref{fig6}, including the $3/4$-filled case for the Co type-III.
	
	These results clearly indicate that the inclusion of both SOC and $U$ effects results in the opening of a band gap in the $j_{\rm{eff}}=1/2$ bands at half filling in type-I and III %\blue{
	heterostructures excluding Mn type-III %}
	and Co type-III. This suggests the formation of spin-orbit coupled Mott insulators in the Ir honeycomb layers, which are cornerstone of the Kitaev candidate materials~\cite{Jackeli2009}, motivating us to further investigate the effective exchange interaction in Sec.~\ref{sec:IV}. %\blue{
	The Co type-III is in an interesting state with $3/4$ filling of $j_{\rm eff}=1/2$ bands, but we exclude it from the following analysis of the effective exchange interactions in Sec.~\ref{sec:IV}. 
	
	%\red{[we need to mention about the exception of Co type III]} \blue{[I also exclude the case of Mn type-III in exchange interactions.]}
	%\blue{[Do we need to list the transfer integrals for type-I and III?]} \red{[Maybe this will be too many tables. Instead, can we mention the typical values and the comparison with the band gap?]} \blue{[For $A$ = Ni in type-III, it holds the largest band gap $\sim 0.484$~eV, but the corresponding largest value of transfer integral is $\sim 0.858$~eV, which is larger than band gap. I also checked all the data, the transfer integrals are all larger than band gap. Besides, I rechecked the data in Jang's paper, the band gap for bulk system is $\sim 0.4$~eV, but the largest transfer integral is $\sim 1.45$~eV.]} \red{[That is the reason why we introduced $U_p$; without $U_p$ the estimates of $K$ are comparable to the band gap, which is not appropriate for the perturbation. Btw, I notice that the estimates for bulk in Fig.~7 seem different from Jang-kuns'. Why?]} \blue{[In Jand-san's work, the transfer integrals are calculated by LDA+SOC. However in Fig.~7, I estimated the coupling constants  by LDA only. If I use the transfer integral from LDA+SOC calculations, they are the same.]}
	
	%\blue{
	Distinct from the emergence of spin-orbit coupled Mott insulator, %} 
	the LDA+SOC+$U$ band structures show metallic states for type-II heterostructures. The $j_{\rm{eff}}=1/2$ bands do not show a clear gap and cross the Fermi level, resulting in the spin-orbit coupled metals. Notably, in all cases, the upper $j_{\rm{eff}}=1/2$ bands are partially doped, realizing electron-doped Mott insulators. The doping rate varies with $A$ atoms. We note that the type-III heterostructure of Mn also exhibits a metallic state, but in this case, holes are doped to the lower $j_{\rm eff}=1/2$ band. See Appendix~\ref{appendix:B} for 
	the orbital projected band structures. 

	%\blue{
	We summarize the electronic states in Table~\ref{tabIII}. 
	The type-I and III heterosuructures are all spin-orbit coupled Mott insulators (SOCI) except for the type-III Mn case, while the type-II are all spin-orbit coupled metals (SOCM). 
	For the SOCM, we also indicate the nature of carriers, electrons or holes; the type-II heterostructures are all electron doped, while the type-III Mn is hole doped.
	%For simplicity, we list the states of each heterostructure, summarized in Table \ref{tabIII}, in which SOCI and SOCM denote spin-orbit coupled insulator and metal. For type-II heterostructures, the metallic states originate from the electron doping, while for Mn type-III, it comes from the hole doping. %}

	%\red{[We may add a summary table for the states of each heterostructure, such as SOCI and SOCM with hole/electron doping. (SOCI(M) = spin-orbit coupled insulator (metal))]} 
	
		\begin{table}
		\caption{\label{tabIII}
			Electronic states of each heterostructure obtained by the LDA+SOC+$U$ calculations. SOCI and SOCM denote spin-orbit coupled insulator and metal, respectively. e and h in the parentheses represent the carriers in the SOCM doped to the mother SOCI. The asterisk for the Co type-III indicates that the system is in the $3/4$-filled insulating state of the $j_{\rm eff}=1/2$ bands.}
		\begin{ruledtabular}
			\begin{tabular}{cccc%c
				}
				%&
				$A$&type-I&II&III \\
				\hline
				%&
				Mn&SOCI&SOCM%-electron
				(e) &SOCM%-hole
				(h)\\
				%&
				Fe&SOCI&SOCM%-electron
				(e)&SOCI \\
				%&
				Co&SOCI&SOCM%-electron
				(e)&SOCI* \\
				%&
				Ni&SOCI&SOCM%-electron
				(e)&SOCI \\
			\end{tabular}			
		\end{ruledtabular}
	\end{table}

	%\red{[Do you have any idea for the reason why the type-II become metallic?]} \blue{[In my opinion, maybe the emergence of two-dimensional electron gas leads to the metallic properties in type-II heterostructure. The unbalanced chemical valence of Ti and Mg atoms accumulates and reconstructs the electrons in the interface, which could only move in the $xy$ plane and result in the formation of 2DEG.]} \red{[That sounds interesting. Can you support it, e.g., by calculating charge distributions? (if it is hard, forget about this)]} \blue{[It is not so difficult, I will add the charge distributions along $z$-axis in the next version.]}
	
	\section{\label{sec:IV}EXCHANGE INTERACTIONS}
\begin{figure}
		\includegraphics[width=1.0\columnwidth]{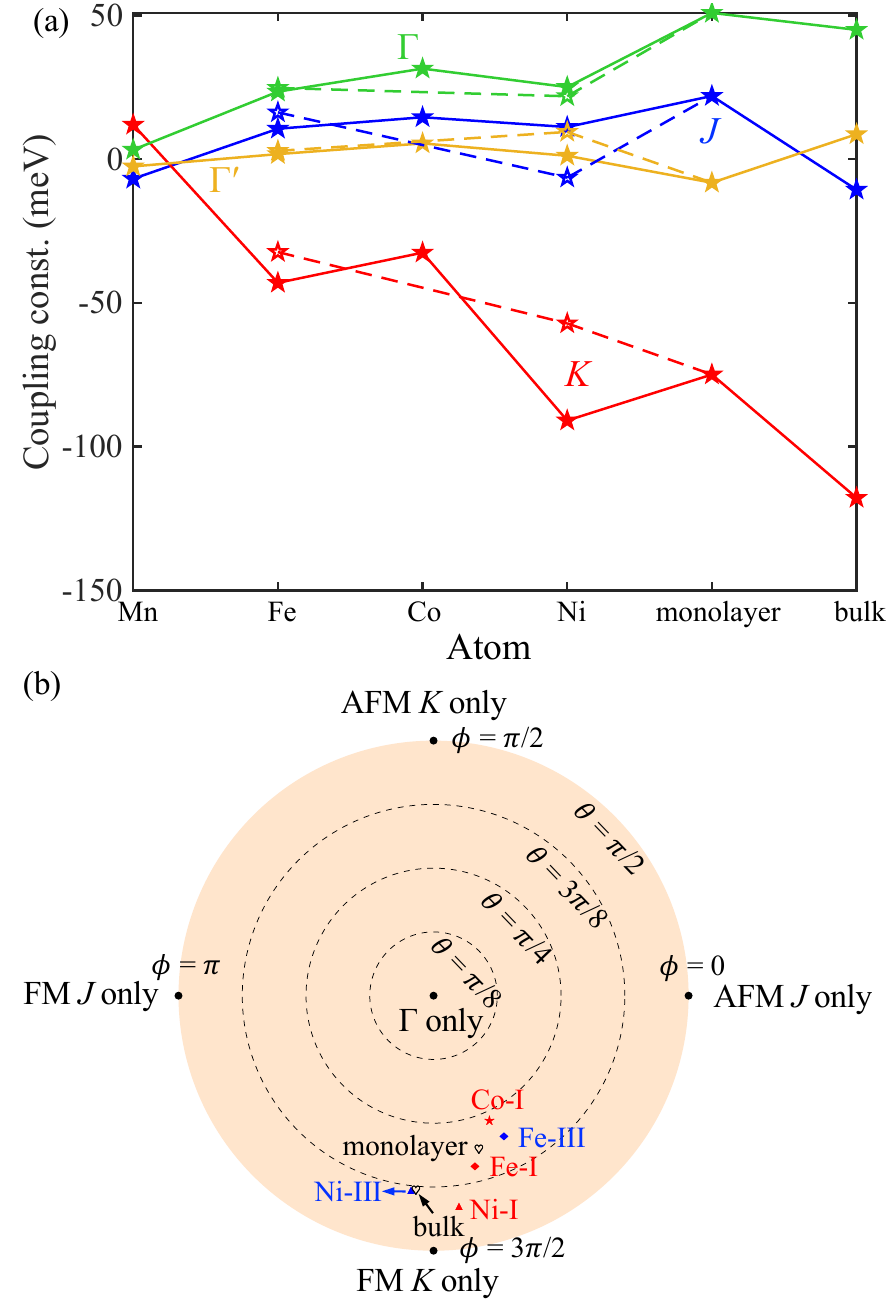}
		\caption{\label{fig7}The effective magnetic constants of heterostructures for different $A$ atoms ($A=$ Mn, Fe, Co, and Ni) of (a) type-I %\blue{
		(solid line with pentagram) %} 
		and type-III %\blue{
		(dashed line with hollow pentagram)%}
		. For comparison, we also show the results for monolayer and bulk. In (b), we summarize the results of $K$, $J$, and $\Gamma$ in (a) except for Mn type-I by using the parametrization in Eq.~\eqref{eq:KJG}. The parameters of the intraorbital Coulomb interaction, Hund's coupling, and spin-orbit coupling are set to $U_{\rm Ir}=3.0$~eV, $J_{\rm H}$/$U_{\rm Ir}=0.1$, and $\lambda= 0.4$~eV, respectively, in the perturbation calculations.} %\red{[may we use black for bulk and monolayer?]} \red{[it would be helpful to indicate $\theta$ and $\phi$ like in Ref.~\cite{Rau.2014}]}}
	\end{figure}

	The electronic band structure analysis reveals the emergence of the spin-orbit coupled Mott insulating state in the MgIrO$_6$ layer of type-I and III heterostructures%\blue{
	, except for the Mn and Co type-III%}
	. %\red{[may we omit Co type III from Fig.~7?]}
	In these cases, the low-energy physics is expected to be described by effective pseudospin models with dominant Kitaev-type interactions~\cite{Jackeli2009}. The effective exchange interactions can be derived by means of the second-order perturbation for the multiorbital Hubbard model (Secs.~\ref{subsec:IIC} and \ref{subsec:IID}). We set $U_{\rm Ir}=3.0$~eV, $J_{\rm H}$/$U_{\rm Ir}=0.1$, and $\lambda= 0.4$~eV in the perturbation calculations. %The pseudospin Hamiltonian is written in the form of 
	%\begin{equation}
		%H %_{\rm{K}}$ 
		%= %$
		%\sum\limits_{\gamma = x,y,z} \sum\limits_{\left\langle i,j\right\rangle}\bm{S}_i^{\rm T} %\bm{J}_{ij}^{\gamma} \bm{S}_j%$
		%,
	%\end{equation}
	%where $i,j$ denote the neighboring sites, and $\gamma$ denotes the three types of Ir-Ir bonds on the MgIrO$_6$ honeycomb layer.%As mentioned above \red{[where?]}, 
	%In our study, the $z$ bond, denoted as $\bm J_{ij}^{z}$, is expressed in term of matrix as follow 
	%\begin{equation}
		%\bm J_{ij}^{z} =
		%\begin{bmatrix}
			%J & \Gamma &\Gamma^\prime \\
			%\Gamma & J & \Gamma^\prime \\
			%\Gamma^\prime & \Gamma^\prime & K \\
		%\end{bmatrix}
	%\end{equation}
	%where the coupling constants for $J$, $K$, $\Gamma$, and $\Gamma^\prime$ represent the isotropic Heisenberg interaction, the bond-dependent Ising-like Kitaev interaction, and two types of the symmetric off-diagonal interactions. \red{[Eqs.~(12) and (13) would be better included in Sec.~II~D.]}
	%We estimate the coupling constants for the MgIrO$_6$ layer for type-I and type-III heterostructures with $A$ = Mn, Fe, Co, and Ni. The parameters of $U%$ = 3$~eV, $\lambda%$ = 0.4$~eV, and $J_{\rm{H}}%$/%$U%$ = 0.1$ for Ir atoms are used in our work~%\blue{\cite{Yamaji2014}%}. \red{[This info would also be included in Sec.~II~D?]} 
	The results are summarized in Fig.~\ref{fig7}. For comparison, we also plot the estimates for monolayer and bulk MgIrO$_3$%\blue{
	. For the bulk case, its potential for hosting Kitaev spin liquids was demonstrated in the previous study~\cite{Jang2021}. Regarding the monolayer, we obtain the results from the band structures shown in Appendix~\ref{appendix:C}, which illustrate the preservation of the $j_{\rm{eff}}=1/2$ manifold and spin-orbit coupled insulating nature.  
	In type-I heterostructures, the dominant interaction is the FM Kitaev interaction $K<0$ for almost all $A$ atoms, except for Mn. 
	%\red{[do not omit Mn type-I from Fig.~\ref{fig7}(a)]} 
	Particularly for $A$ = Ni, the absolute value of $K$ is significantly larger than the others, even considerable when compared with the monolayer and bulk MgIrO$_3$. The subdominant interaction is the off-diagonal symmetric interaction $\Gamma>0$. The other off-diagonal symmetric interaction $\Gamma'$ as well as the Heisenberg interaction $J$ is weaker than them. In the Mn case, all the interactions are exceptionally weak, presumably because of the intervening Mn $3d$ band and its hybridization with the Ir $j_{\rm eff}=1/2$ bands. 
	%\red{[We may mention the possible reason why the Mn case is exceptional.]} \blue{[I am not sure of the reason for this phenomenon. Is this because of the Mn-3d bands residing between upper and lower $j_{\rm eff}=1/2$ bands?]} \red{[It occurs also for the Ni case. Do you see any particular orbital hybridization for the Mn case?]} \blue{[For Mn type-I, the inserted bands between upper and lower $j_{\rm eff}=1/2$ bands are dominant by Mn $3d$ and O $2p$ orbitals, with slight hybridization with Ir $5d$ orbitals. But for Ni type-I, the pure hybridization of Ni $3d$ and O $2p$  predominate in the inserted bands, without Ir $5d$ orbitals.]}
	Meanwhile, for the type-III %\blue{
	heterostructures%}
	, since the Mn and Co cases exhibit a metallic state and $3/4$ occupation of $j_{\rm eff}=1/2$ bands, respectively, we only calculate the effective magnetic constants for Fe and Ni. In these cases also, the dominant interaction is the FM $K$, accompanied by the subdominant $\Gamma$ interaction, as shown in Fig.~\ref{fig7}(a).  %}
	
	%\blue{[The largest value of transfer integral for heterostructures is about 1041 meV for type-I with $A$ = Ni], which is much smaller than 1450 meV in Jang-san's paper}
	
	Thus, in all cases except the Mn type-I heterostructure, the dominant magnetic interaction in the spin-orbit coupled Mott insulating state in the Ir honeycomb layer is effectively described by the FM Kitaev interaction. 
	Since $\Gamma^\prime$ is smaller than the other exchange constants, the low-energy magnetic properties can be well described by the generic %\blue{
	$K$-$J$-$\Gamma$ %} 
	model~\cite{Rau.2014,Rusna2019}, which has been widely and successfully applied to study the Kitaev QSLs. We summarize the obtained effective exchange interactions %\blue{
	of $K$, $J$, and $\Gamma$ %with angle $\theta$ and $\phi$  
	by using the parametrization 
	\begin{equation}
		(K, J, \Gamma) = \mathcal{N}(\rm{sin}\theta \rm{sin}\phi, \rm sin\theta \rm cos\phi, \rm cos\theta),
		\label{eq:KJG}
	\end{equation}
	where $\mathcal{N} = (K^2+J^2+\Gamma^2)^{-1/2}$ is the normalization factor. Figure~\ref{fig7}(b) presents the  results except for Mn type-I. Our heterostructures distribute in the region near the FM $K$ only case ($\theta=\pi/2$ and $\phi=3\pi/2$). 
	%\red{[Let's omit the Mn-I from Fig.~7(c).]} \blue{[I have deleted the data of Mn in Fig. 7(a) and 7(b).]}
	We find a general trend that larger $A$ atoms make the systems closer to the FM $K$ only case; the best is found for Ni type-I and III. %\blue{[In comparison, we then map the location of the effective exchange interaction in the phase diagram of the model with $\Gamma >$ 0, as previously discussed in~\cite{Rau.2014}. In the classical phase diagram, we find the majority of the exchange coupling constants lie within the incommensurate spiral phase. However, exceptions are found for $A$ = Mn (Ni) in type-I (III) and for $A$ = Mn in type-III, which exhibit zigzag and AFM states, respectively. While in the exact diagonalization (ED) phase, most of the ground states are spiral phase, except for $A$ = Mn in type-I (III) which is zigzag and AFM state, respectively.]} 
	In the previous studies for the $K$-$J$-$\Gamma$ model~\cite{Rau.2014,Rusna2019}, a keen competition between different magnetic phases was found in this region, which does not allow one to conclude the stable ground state in the thermodynamic limit. Given that this region appears to be connected to the solvable point for the FM Kitaev QSL, our heterostructures provide a promising platform for investigating the Kitaev QSL physics and related phase competition by finely tuning the magnetic interactions via the proximity effect in the heterostructures.

	\section{\label{sec:VI}DISCUSSION}

	%\blue{
	Our systematic study of ilmenite heterostructures MgIrO$_3$/$A$TiO$_3$ with $A=$ Mn, Fe, Co, and Ni reveals their fascinating electronic and magnetic properties. The heterostructures in the paramagnetic state are metallic in terms of band structures obtained by LDA+SOC, regardless of types and $A$ atoms. When incorporating the effect of electron correlation by the LDA+SOC+$U$ calculations, type-II heterostructures remain metallic across entire $A$ atoms, whereas type-I and III heterostrcutures turn into insulating states, except for Mn type-III. As a consequence, the electronic states of heterostructures are classified into the spin-orbit coupled insulators and metals, each holding unique properties. The insulating cases possess the $j_{\rm eff}=1/2$ pseudospin degree of freedom, and furnish a fertile playground to investigate the Kitaev QSL. In these cases, however, due to the magnetic proximity effects from the $A$ layer, we may expect interesting modification of the QSL state, as discussed in Sec.~\ref{subsec:VIA} below. Meanwhile, the metallic cases open avenues for exploring spin-orbit coupled metals, relatively scarce in strongly correlated systems~%\blue{
	\cite{hanawa2001,ohgushi2011,harter2017parity,hiroi2018pyrochlore,tajima2020spin}%}
	. %}
	%\red{[For both compounds, please try to cite more original ones. For Cd$_2$Re$_2$O$_7$, check the review in JPSJ 87, 024702 (2018). For Pb$_2$Re$_2$O$_7$, see PRB 83, 125103 (2011).]} 
	In Sec.~\ref{subsec:VIB}, we discuss the possibility of exotic superconductivity in our self-doped heterostructures. In addition, we discuss the feasibility of fabrication of these heterostructures and identification of the Kitaev QSL nature in experiments in Sec.~\ref{subsec:VIC}.
	
	%\subsection{ \label{subsec:VIA}Access to superconductor}
	\subsection{ \label{subsec:VIA}Majorana Fermi surface by magnetic proximity effect}
	In the pure Kitaev model, the spins are fractionalized into itinerant Majorana fermions and localized $Z_2$ gauge fluxes~\cite{kitaev2006anyons}. The former has gapless excitations at the nodal points of the Dirac-like dispersions at the K and K' points on the Brillouin zone edges, while the latter is gapped with no dispersion. When an external magnetic field is applied, the Dirac-like nodes of Majorana fermions are gapped out, resulting in the emergence of quasiparticles obeying non-Abelian statistics~\cite{kitaev2006anyons}. Beyond the uniform magnetic field, the Majorana dispersions are  further modulated by an electric field and a staggered magnetic field~\cite{chari2021,Nakazawa2022}. For instance, with the existence of the staggered magnetic field, the Dirac-like nodes at the K and K' points are shifted in the opposite directions in energy to each other, leading to the formation of the Majorana Fermi surfaces. Moreover, the introduction of both uniform and staggered magnetic fields can lead to further distinct modulations of the Majorana Fermi surfaces around the K and K' points, which are manifested by nonreciprocal thermal transport carried by the Majorana fermions~\cite{Nakazawa2022}. 
		
	In our heterostructures of type-I and III, the $A$ layer supports a N\'eel order in most cases (Table~\ref{tab:table2}). It can generate an internal staggered magnetic field applied to the Ir layer through the magnetic proximity effect. This mimics the situations discussed above, and hence, it may result in the Majorana Fermi surfaces in the possible Kitaev QSL in the Ir layer. The combination of the uniform and staggered magnetic fields could also be realized by applying an external magnetic field to these heterostructures.
	Thus, the ilmenite heterostructures in proximity to the Kitaev QSL in the Ir layer hold promise for the formation of Majorana Fermi surfaces and resultant exotic thermal transport phenomena, providing a unique platform for identifying the fractional excitations in the Kitaev QSL. 
	
	\subsection{ \label{subsec:VIB}Exotic superconductivity by carrier doping}
	QSLs have long been discussed as mother states of exotic superconductivity~\cite{anderson1987resonating,lee2006doping,broholm2020quantum}. % \red{[Is this an over statement? Is there experimental evidence that the doped QSL shows a superconductivity? As far as I know, it is an expectation or an interpretation of high-$T_c$ superconductivity that  carrier doping to QSL could give rise to superconductivity.]} 
	%\red{[You may also touch on the possible exotic superconductivity in a doped SOCI in Ir 214 compounds; see for instance PRL 110, 027002.]}
	There, the introduction of mobile carriers to insulating QSLs possibly induces superconductivity in which the Cooper pairs are mediated by strong spin entanglement in the QSLs. A representative example discussed for a long time is high-$T_c$ cuprates; here, the $d$-wave superconductivity is induced by carrier doping to the undoped antiferromagnetic state that is close to a QSL of so-called resonating valence bond (RVB) type~%\blue{
	\cite{bednorz1986possible,%cava1987,tarascon1987,
	anderson1987,%blumberg1997evolution,yang2008emergence,
	lee2006doping}%}
	. A similar exotic superconducting state was also discussed for an iridium oxide Sr$_2$IrO$_4$ with spin-orbital entangled $j_{\rm eff}=1/2$ bands~\cite{watanabe2013,yan2015}. 
	Carrier doped Kitaev QSLs have also garnered extensive attention due to its potential accessibility to unconventional superconductivity that may possess more intricate paring from the unique QSL properties. It was reported that doping into the Kitaev model with additional Heisenberg interactions ($K$-$J$ model) led to a spin-triplet %\blue{
	topological %} 
	superconducting state~\cite{You2012}, where the pairing nature  is contingent upon the doping concentration. Furthermore, the competition between $K$ and $J$ also significantly impacts the superconducting state; for example, $K$ prefers a $p$-wave superconducting state, whereas $J$ tends to favor a $d$-wave one~\cite{Hyart2012,Okamoto2013}. Even topological superconductivity is observed in an extended $K$-$J$-$\Gamma$ model~\cite{schmidt2018}.  
	
	In the present work, we found metallic states in the spin-orbital coupled $j_{\rm eff}=1/2$ bands in the Ir layer for all type-II heterostructures (electron doping) and the type-III Mn heterostructure (hole doping) (see Table~\ref{tabIII}). Besides, in the type-III Co heterostructure, electron doping occurs in the Ir layer, resulting in the $3/4$-filled insulating state in the $j_{\rm eff}=1/2$ bands. These appealing results suggest that our ilmenite heterostructures offer a platform for studying exotic metallic and superconducting %\blue{
	(even topological) %} 
	properties with great flexibility by various choices of materials combination, which have been scarcely realized in the bulk systems. %}

	\subsection{\label{subsec:VIC}Experimental feasibility}
	The bulk compounds of ilmenite $A$TiO$_3$ with $A$ = Mn, Fe, Co, and Ni have been successfully synthesized and investigated for over half a century due to its fruitful magnetic and novel electronic properties~\cite{ishikawa1957magnetic,newnham1964crystal,shirane1959neutron,kato1982metamagnetic,Yuan2020}. Technically, the Fe case, however, is more challenging compared to the others, as its synthesis needs very high pressure and high temperature %~\cite{raghavender2013nano,wechsler1984crystal} 
	conditions~\cite{wechsler1984crystal,raghavender2013nano}. Besides, the iridium ilmenite MgIrO$_3$ has also been synthesized as a power sample%~\cite{Haraguchi.2018,negishi2023composition}
	, where a magnetic phase transition was observed at $31.8$~K~\cite{Haraguchi.2018}. The experimental lattice parameters are $5.14$~\AA~for $A$TiO$_3$ with $A$ = Mn~%\blue{
	\cite{shirane1959neutron,yamauchi1983spin,ko1988high,Mufti2011}%}
	, %\red{[why don't you cite more original papers, like J. Magn. Magn. Mater. 31-34, 1071 and Phys. Chem. Miner. 15, 355 (1988)? The same applies to the followings.]}
	 $5.09$~\AA~for $A$ = Fe~%\blue{
	\cite{ishikawa1956study,ishikawa1958magnetic,wechsler1984crystal}%}
	, $5.06$~\AA~for $A$ = Co~%\blue{
	\cite{newnham1964crystal,acharya2016structural,yuan2020dirac}%}
	, and $5.03$~\AA~for $A$ = Ni~%\blue{
	\cite{ishikawa1958magnetic,shirane1959neutron,heller1963antiferromagnetism,Harada2016}%}
	, respectively, as well as that is $5.16$~\AA~for MgIrO$_3$. The relatively small lattice mismatch between these materials also ensure the possibility of combining them to create heterostructure with different compounds. 
	Indeed, we demonstrated this in Sec.~\ref{sec:IIa}; see Table~\ref{tab:table1}. 
	More excitingly, the IrO$_6$ honeycomb lattice has been successfully incorporated into the ilmenite MnTiO$_3$ with the formation of several Mn-Ir-O layers~\cite{miura2020stabilization}. This development lightens the fabrication of a supercell between MgIrO$_3$ and $A$TiO$_3$. %~\cite{miura2020stabilization}. 
	
	%\red{[I believe the inelastic neutron scattering is impossible for heterostructure samoles. Instead, we may suggest the Raman scattering as a good probe of fractional Majorana excitations. You should cite the previous studies for the bulk and the atomically thin films.]} 
	The verification of Kitaev QSL poses a significant challenge even though the successful synthesis of aforementioned heterostructures. First of all, it is crucial to identify the spin-orbital entangled electronic states with the formation of the $j_{\rm eff}=1/2$ bands in these heterostructures, as they are essential for the Kitaev interactions between the pseudospins. Several detectable spectroscopic techniques are useful for this purpose, applicable to both bulk and heterostructures~%\blue{
	\cite{Kim2008,comin2012,sohn2013,gretarsson2013,Plumb2014,koitzsch2016,zhou2016,sinn2016electronic,suzuki2021proximate}%}
	. %\blue{
	Even the Kitaev exchange interaction can be directly uncovered in experiment~\cite{hwan2015direct,das2019}. %} %\red{[Are these references appropriate? Please try to cite the original relevant papers as much as possible. Also, I think not only X-ray but also, for instance, ARPES.]} 
	%red{[Not only the $j_{\rm eff}=1/2$ bands but also the Kitaev-type interactions have also been identified experimentally; for instance, see Nat. Phys. 11, 462 (2015) and Phys. Rev. B 99, 081101(R) (2019).]} 
	However, the key challenge lies in probing the intrinsic properties of Kitaev QSL, such as fractional spin excitations. Thus far, despite cooperative studies between theories and experiments on, for instance, dynamical spin structure factors~%\blue{
	\cite{knolle2014,knolle2015,banerjee2016proximate,yoshitaka2016,do2017majorana,yoshitake2017,yoshitake20172} %} %\red{[cite also theory papers]}
	%\red{[cite also PRL 112, 207203 (2014), PRB 92, 115127 (2015), PRB 96, 024438 (2017), and PRB 96, 064433 (2017). omit song2016 and nasu2021.]} 
	 and the thermal Hall effect and its half quantization~%\blue{
	\cite{nasu2017,kasahara2018majorana,yokoi2021half,chern2021}%}
	, 
	 %\red{[cite also theory papers]}, 
	%\red{[omit nasu2020]} 
	 have been developed to identify the fractional excitations in Kitaev QSL, directly applying them on the heterostructures is still a great challenge. A promising experimental tool would be the Raman spectroscopy, given its successful application to not only bulk~\cite{sandilands2015,nasu2016fermionic,glamazda2016raman} but also atomically thin layers~%\blue{
	\cite{zhou2019possible,lee2021multiple}%}
	. %\red{[cite also J. Phys. Chem. Solids 128, 291 (2019).]} %\cite{nasu2016fermionic,sandilands2015,glamazda2016raman,lee2021multiple}. 
	%\red{[We may also mention about many theoretical proposals on local probes like STM.]}
	The signals might be enhanced by piling up the heterostructures. 
	%\blue{
	Besides, many proposals for probing the Kitaev QSL in thin films and heterostructures have been recently made, such as local probes like scanning tunneling microscopy (STM) and atomic force microscopy (AFM) \cite{feldmeier2020,pereira2020,konig2020,udagawa2021,bauer2023} as well as the spin Seebeck effect \cite{kato2024spin}. %}
	Additionally, as mentioned in Sec.~\ref{subsec:VIA}, the observation of the Majorana Fermi surfaces by thermal transport measurements in some particular heterostructures is also interesting. 
	%\red{[You may also touch on, for instance, the spectroscopy measurements to confirm the spin-orbital entangled electronic state, thermal transport measurement for the Majorana Fermi surface effects, and spin Seebeck effect discussed by Kato-san recently.]} 
	%\redout{Our ilmenite heterostructures would offer a distinctive playground for verifying the Kitaev QSL.}
	
	%\red{[This is for the insulating Kitaev QSLs. So might be better to bring this section before the discussion on the carrier doping in type-II. Please think about the reorganization of the two sections.]}  

	\section{\label{Sec:VII}Summary}
	To summarize, we have conducted a systematic investigation of the electronic and magnetic properties of the bilayer structures composed by the ilmenites $A$TiO$_3$ with $A$ = Mn, Fe, Co, and Ni, in combination with the candidate for Kitaev magnets MgIrO$_3$. We have designed and labeled three types of heterostructure, denoted as type-I, II, and III, distinguished by the atomic configurations at the interface. Our analysis of the electronic band structures based on the {\it ab initio} calculations has revealed that the spin-orbital coupled bands characterized by the pseudospin $j_{\rm eff}=1/2$, one of the fundamental component for the Kitaev interactions, is retained in the MgIrO$_3$ layer for all the types of heterostructures. We found that the MgIrO$_3$/$A$TiO$_3$ heterostructures of type-I and III are mostly spin-orbit coupled insulators, while those of type-II are spin-orbit coupled metals, irrespective of the $A$ atoms. In the insulating heterostructures of type-I and III, based on the construction of the multiorbital Hubbard models and the second-order perturbation theory, we further found that the low-energy magnetic properties can be described by the $j_{\rm eff}=1/2$ pseudospin models in which the estimated exchange interactions are dominated by the Kitaev-type interaction. We showed that the parasitic subdominant interactions depend on the type of the heterostructure as well as the $A$ atoms, offering the playground for systematic studies of the Kitaev spin liquid behaviors. Moreover, the stable N\'eel order in the $A$TiO$_3$ layer acts as a staggered magnetic field through the magnetic proximity effect, leading to the potential realization of Majorana Fermi surfaces in the MgIrO$_3$ layer. %\blue{
	Meanwhile, in the metallic heterostructures of type-II as well as type-I Mn, we found that the nature of carriers and the doping rates vary depending on the heterostructures. This provides the possibility of systematically studying the spin-orbit coupled metals, including exploration of unconventional superconductivity due to the unique spin-orbital entanglement. %Moreover, we discussed the potential realization of exotic properties, including the emergence of Majorana Fermi surface through magnetic proximity effect in insulating heterostructures, and carrier-doping induced superconductivity in metallic heterostructures. Moving forward, we anticipate the possible realization of these heterostructures and the development of detectable measurements to explore their unique properties further. %} %\red{[summarize the results for the metallic cases]} \red{[also wrap up the discussion sections]}
	
	%\blue{
	In recent decades, significant progress has been made in the study of QSLs, primarily focusing on the discovery and expansion of new members in bulk materials. However, there has been limited exploration of creating and manipulating the QSLs in heterostructures despite the importance for device applications. Our study has demonstrated that the Kitaev-type QSL could be surveyed in ilmenite oxide heterostructures, displaying remarkable properties distinct from the bulk counterpart, such as flexible tuning of the Kitaev-type interactions and other parasitic interactions, and carrier doping to the Kitaev QSL. %are tunable by external fields such as magnetic and ferroelectric polarization fields. In monolayer $\alpha$-RuH$_{3/2}X_{3/2}$ ($X=$ Cl and Br), for instance, antiferromagnetic Kitaev interactions are realized due to symmetry breaking by a polar field, suggesting that tunable Kitaev interactions triggered by polarization flip is possible. The polarization flip not only modulates the interactions but also affects the Fermi level, offering continuous control over the doping rates, which may influence the emergence of superconductivity. On the other hand, 
	Besides the van der Waals heterostructures such as the combination of $\alpha$-RuCl$_3$ and graphene, our finding would enlighten an additional route to explore the Kitaev QSL physics including the utilization of Majorana and anyonic excitations for future topological computing devices. %In principle, it is easier to achieve in experiments than oxide heterostructures, mainly due to the weaker interaction than the oxidized covalent bond. %}
	
	%\blue{[I am not sure whether these expressions are appropriate.]}
	%Our findings indicate the Kitaev-type interactions could not only be realized in two-dimensional heterostructural systems, such as graphene/$\alpha$-RuCl$_3$, but also surveyed in three-dimensional heterostructures. \red{[I feel this sentence is abrupt and might mislead the readers]} We also hope these results can be validated in experiments. \red{[for PRX, maybe we need more perspectives]} 

	\begin{acknowledgments}
		
		We thank Y. Kato, M. Negishi, S. Okumura, A. Tsukazaki, and L. Zh. Zhang, for fruitful discussions. This work was supported by JST CREST Grant (No.~JP-MJCR18T2). Parts of the numerical calculations were performed in the supercomputing systems of the Institute for Solid State Physics, the University of Tokyo.

	\end{acknowledgments}
	
		\appendix
		
		\section{\label{appendix:A}Detailed {\it ab initio} data for energy and magnetic coupling}
		
		In this Appendix, we present the details of {\it ab initio} results for various types of heterostructures. Tables~\ref{tab:s1}-\ref{tab:s4} list the energy differences between different magnetic states for MgIrO$_3$/$A$TiO$_3$ heterostructures with $A$ = Mn, Fe, Co, and Ni. The bold elements in these tables are the lowest-energy state in each type, utilized for the calculations of band structures in Sec.~\ref{subsubIIIB:1}. We also show in Table~\ref{tab:s5} the effective magnetic coupling constants between $A$ atoms, in which negative and positive value indicates the FM and AFM coupling, respectively. Note that the $A$ atoms comprise a triangular lattice at the interface in type I, a honeycomb lattice at the $A$TiO$_3$ layer, and a honeycomb lattice at the interface, as depicted in Fig.~\ref{fig1}. 
		%\red{[I am wondering if the data in Table~\ref{tab:s5} are consistent with those in Tables~\ref{tab:s1}-\ref{tab:s4}. For example, the magnetic coupling constant for type III Co is very large $\sim 6.5$~eV, but the energy difference between FM and N\'eel in Table~\ref{tab:s3} is quite small, less than $1$~meV.]} \blue{[Sorry for my mistake, I have checked the unit that should be meV keeping in line with coupling constants in Fig.~\ref{fig7}, rather than eV.]}
		
		\begin{table}
			\caption{\label{tab:s1}
				Energy differences between different magnetic ordered states obtained by the LDA+SOC+$U$ calculations for MgIrO$_3$/MnTiO$_3$: FM, N\'eel, and zigzag denotes the ferromagnetic, N\'eel-type antiferromagnetic, and zigzag-type antiferromagnetic orders, respectively. While the directions of the magnetic moments are all in-plane for the Ir layers, those for $A$ can be in-plane (``in") or out-of-plane (``out"). The bold numbers denote the low-energy states used for calculating the band structures in Sec.~\ref{subsubIIIB:1}. 
			}
				\begin{ruledtabular}
			\begin{tabular}{cccccc%c
				}
				%&
				\multicolumn{3}{c}{magnetic state}&\multicolumn{3}{c}{energy/Ir (meV)} \\
				%&
				Ir&\multicolumn{2}{c}{Mn}&I&II&III \\
				\hline
				%&
				\multirow{4}{*}{in-FM}&\multirow{2}{*}{in}&FM&323.5&4.868&975.4 \\
				%&~
				&~&N\'eel&1.492&46.52&1002 \\
				%&~
				&\multirow{2}{*}{out}&FM&18.13&4.268&1065 \\
				%&~
				&~&N\'eel&3.695&44.554&978.2 \\
				\hline
				%&
				\multirow{4}{*}{out-FM}&\multirow{2}{*}{in}&FM&4.266&7.343&990.8 \\
				%&~
				&~&N\'eel&0.773&43.68&993.2 \\
				%&~
				&\multirow{2}{*}{out}&FM&35.71&1.584&1014 \\
				%&~
				&~&N\'eel&76.55&40.34&879.6 \\
				\hline
				%&
				\multirow{4}{*}{N\'eel}&\multirow{2}{*}{in}&FM&27.18&8.127&78.52 \\
				%&~
				&~&N\'eel&\textbf{26.98}&42.83&\textbf{48.51} \\
				%&~
				&\multirow{2}{*}{out}&FM&20.45&\textbf{4.474}&121.0 \\
				%&~
				&~&N\'eel&2.953&39.39&22.45 \\
				\hline
				%&
				\multirow{4}{*}{zigzag}&\multirow{2}{*}{in}&FM&0.307&2.923&78.80 \\
				%&~
				&~&N\'eel&0.000 &21.55&0.000 \\
				%&~
				&\multirow{2}{*}{out}&FM&20.05&0.000&120.49 \\
				%&~
				&~&N\'eel&4.893&39.62&22.95 \\

			\end{tabular}			
		\end{ruledtabular}
		\end{table}
		
				\begin{table}
			\caption{\label{tab:s2}
				Energy differences between different magnetic ordered states for MgIrO$_3$/FeTiO$_3$. The notations are common to Table~\ref{tab:s1}.
			}
			\begin{ruledtabular}
				\begin{tabular}{cccccc%c
					}
					%&
					\multicolumn{3}{c}{magnetic state}&\multicolumn{3}{c}{energy/Ir (meV)} \\
					%&
					Ir&\multicolumn{2}{c}{Fe}&I&II&III \\
					\hline
					%&
					\multirow{4}{*}{in-FM}&\multirow{2}{*}{in}&FM&9.090 &1.437&\textbf{0.000} \\
					%&~
					&~&N\'eel&17.00&15.75&13.14 \\
					%&~
					&\multirow{2}{*}{out}&FM&15.71&21.80&2.631 \\
					%&~
					&~&N\'eel&1.170&27.02&14.25 \\
					\hline
					%&
					\multirow{4}{*}{out-FM}&\multirow{2}{*}{in}&FM&10.68&0.973&0.092 \\
					%&~
					&~&N\'eel&12.12&0.948&13.01 \\
					%&~
					&\multirow{2}{*}{out}&FM&55.58&21.13&1.739 \\
					%&~
					&~&N\'eel&8.115&26.29&8.833 \\
					\hline
					%&
					\multirow{4}{*}{N\'eel}&\multirow{2}{*}{in}&FM&24.64&\textbf{0.000}&0.056 \\
					%&~
					&~&N\'eel&11.80&0.721&12.97 \\
					%&~
					&\multirow{2}{*}{out}&FM&14.48&20.21&1.244 \\
					%&~
					&~&N\'eel&\textbf{0.000}&25.37&8.596 \\
					\hline
					%&
					\multirow{4}{*}{zigzag}&\multirow{2}{*}{in}&FM&15.58&1.233&0.056 \\
					%&~
					&~&N\'eel&4.376 &13.60&4.222 \\
					%&~
					&\multirow{2}{*}{out}&FM&14.35&21.42&2.244 \\
					%&~
					&~&N\'eel&4.743&26.59&9.608 \\
				\end{tabular}			
			\end{ruledtabular}
		\end{table}
		
			\begin{table}
			\caption{\label{tab:s3}
				Energy differences between different magnetic ordered states for MgIrO$_3$/CoTiO$_3$. The notations are common to Table~\ref{tab:s1}.
			}
			\begin{ruledtabular}
				\begin{tabular}{cccccc%c
					}
					%&
					\multicolumn{3}{c}{magnetic state}&\multicolumn{3}{c}{energy/Ir (meV)} \\
					%&
					Ir&\multicolumn{2}{c}{Co}&I&II&III \\
					\hline
					%&
					\multirow{4}{*}{in-FM}&\multirow{2}{*}{in}&FM&188.2 &20.02&0.795 \\
					%&~
					&~&N\'eel&191.3&\textbf{0.000}&0.253 \\
					%&~
					&\multirow{2}{*}{out}&FM&40.23&25.63&72.39 \\
					%&~
					&~&N\'eel&1.201&235.7&87.28 \\
					\hline
					%&
					\multirow{4}{*}{out-FM}&\multirow{2}{*}{in}&FM&0.174&20.06&0.800 \\
					%&~
					&~&N\'eel&\textbf{0.000}&0.019&0.071 \\
					%&~
					&\multirow{2}{*}{out}&FM&295.3&25.85&260.4 \\
					%&~
					&~&N\'eel&47.55&0.044&49.64 \\
					\hline
					%&
					\multirow{4}{*}{N\'eel}&\multirow{2}{*}{in}&FM&189.5&19.56&0.749 \\
					%&~
					&~&N\'eel&184.3&19.36&\textbf{0.000} \\
					%&~
					&\multirow{2}{*}{out}&FM&17.53&24.82&33.16 \\
					%&~
					&~&N\'eel&0.617&234.9&31.16 \\
					\hline
					%&
					\multirow{4}{*}{zigzag}&\multirow{2}{*}{in}&FM&191.8&19.88&0.748 \\
					%&~
					&~&N\'eel&189.6 &19.39&1.969 \\
					%&~
					&\multirow{2}{*}{out}&FM&40.81&25.45&34.02 \\
					%&~
					&~&N\'eel&1.476&130.0&32.03 \\
				\end{tabular}			
			\end{ruledtabular}
		\end{table}

	\begin{table}
	\caption{\label{tab:s4}
		Energy differences between different magnetic ordered states for MgIrO$_3$/NiTiO$_3$. The notations are common to Table~\ref{tab:s1}.
	}
	\begin{ruledtabular}
		\begin{tabular}{cccccc%c
			}
			%&
			\multicolumn{3}{c}{magnetic state}&\multicolumn{3}{c}{energy/Ir (meV)} \\
			%&
			Ir&\multicolumn{2}{c}{Ni}&I&II&III \\
			\hline
			%&
			\multirow{4}{*}{in-FM}&\multirow{2}{*}{in}&FM&2.990 &0.097&37.80 \\
			%&~
			&~&N\'eel&4.998&70.93&35.99 \\
			%&~
			&\multirow{2}{*}{out}&FM&3.536&1.255&26.76 \\
			%&~
			&~&N\'eel&0.156&69.45&35.74 \\
			\hline
			%&
			\multirow{4}{*}{out-FM}&\multirow{2}{*}{in}&FM&6.901&2.039&37.48 \\
			%&~
			&~&N\'eel&7.311&70.74&35.98 \\
			%&~
			&\multirow{2}{*}{out}&FM&20.94&2.385&24.79 \\
			%&~
			&~&N\'eel&6.220&69.61&61.76 \\
			\hline
			%&
			\multirow{4}{*}{N\'eel}&\multirow{2}{*}{in}&FM&14.06&\textbf{1.678}&48.98 \\
			%&~
			&~&N\'eel&13.00&18.68&\textbf{54.35} \\
			%&~
			&\multirow{2}{*}{out}&FM&4.941&0.294&26.51 \\
			%&~
			&~&N\'eel&\textbf{0.000}&18.631&37.36 \\
			\hline
			%&
			\multirow{4}{*}{zigzag}&\multirow{2}{*}{in}&FM&4.971&0.000&36.97 \\
			%&~
			&~&N\'eel&6.275 &33.08&0.000 \\
			%&~
			&\multirow{2}{*}{out}&FM&3.451&0.042&26.20 \\
			%&~
			&~&N\'eel&3.876&68.36&35.45 \\
		\end{tabular}			
	\end{ruledtabular}
\end{table}

	\begin{table}
	\caption{\label{tab:s5}
		Effective magnetic coupling constants between the $A$ atoms for three types of heterostructures. The unit is in meV.}
		\begin{ruledtabular}
	\begin{tabular}{cccc%c
		}
		%&
		$A$&type-I&II&III \\
		\hline
		%&
		Mn&0.667&-0.712&2.417\\
		%&
		Fe&0.017&-0.607&-0.149\\
		%&
		Co&0.065&0.262&6.458\\
		%&
		Ni&0.088&-0.126&0.405\\
	\end{tabular}			
\end{ruledtabular}
\end{table}
		
			\section{\label{appendix:B}Orbital projected band structures}
\begin{figure}[h]
			\includegraphics[width=1.0\columnwidth]{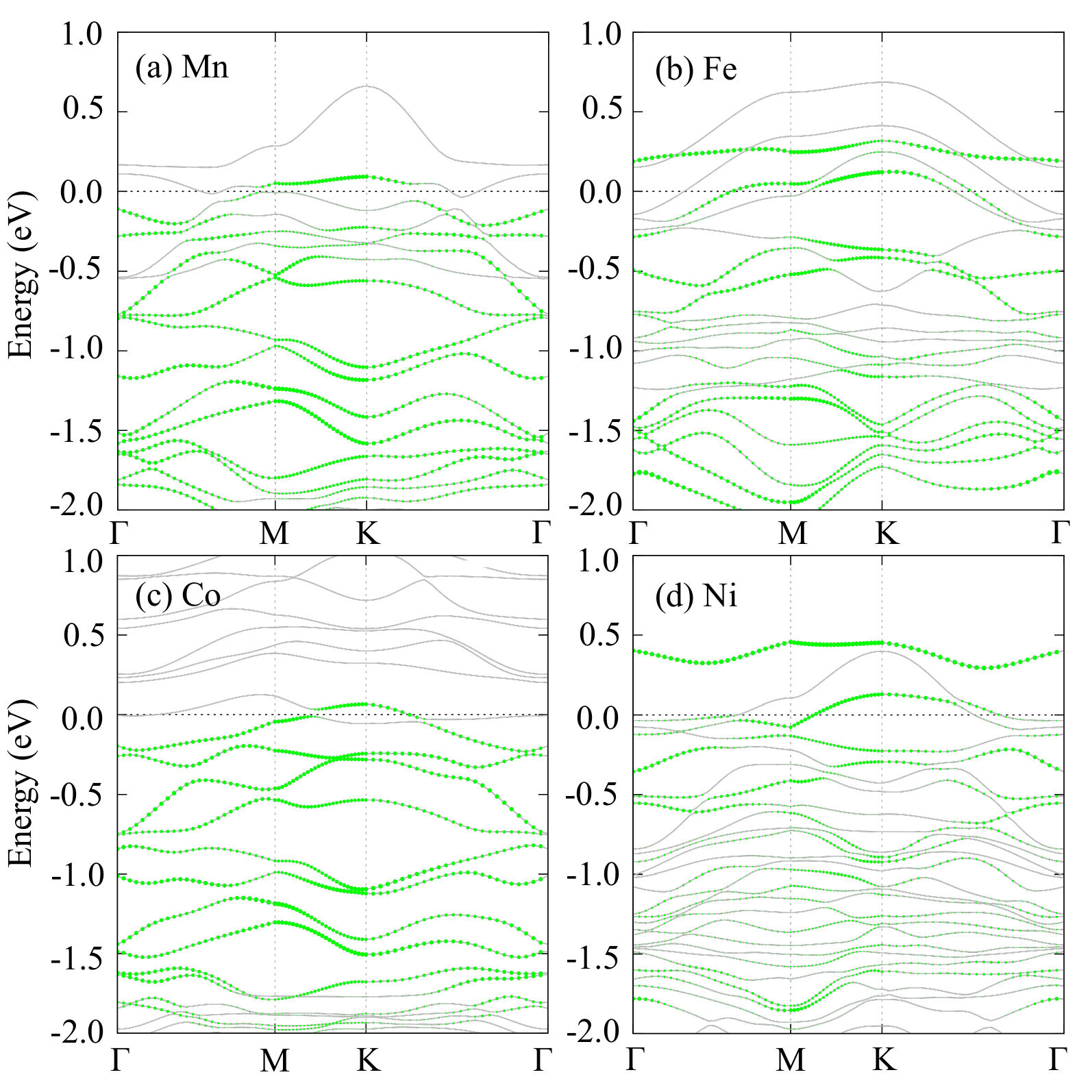}
			\caption{\label{fig8} Projection to the Ir $5d$ orbitals of the band structure for the type-II MgIrO$_3/A$TiO$_3$ heterostructures with (a) $A$ = Mn, (b) Fe, (c) Co, and (d) Ni. The gray lines depict the band structures shown in the middle panels of Figs.~\ref{fig2}(b)-\ref{fig5}(b), and the green shade represents the weight of Ir $5d$ orbitals. The Fermi level is set to zero. %\red{[The characters in the figures are too small (also in Fig.~9)]}
			}
		\end{figure}

\begin{figure}
			\includegraphics[width=1.0\columnwidth]{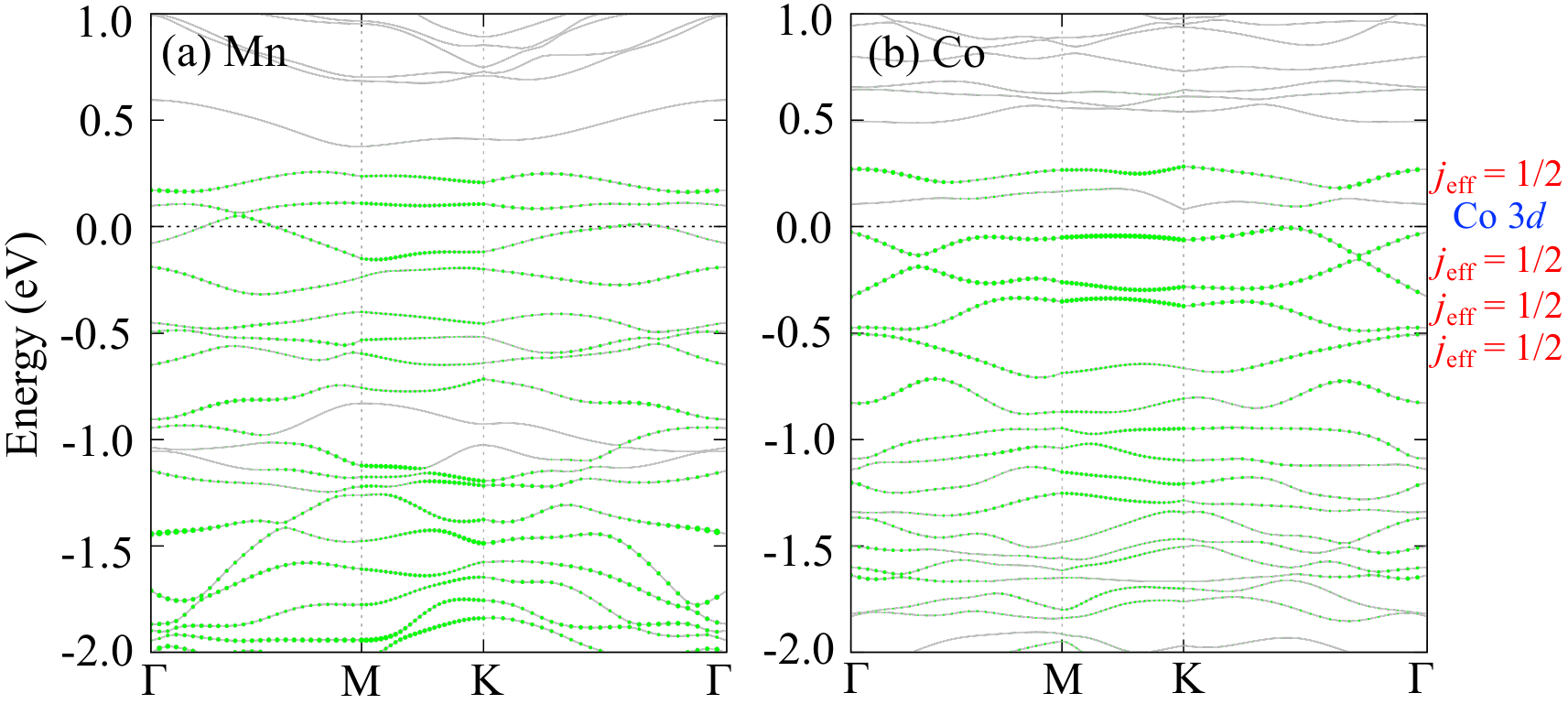}
			\caption{\label{fig9} Projection to the Ir $5d$ orbitals of the band structure for the type-III heterostructures of (a) MgIrO$_3$/MnTiO$_3$ and (b) MgIrO$_3$/CoTiO$_3$.  The notations are common to Fig.~\ref{fig8}. 
			%\red{[align two figures horizontally, as in Fig.~\ref{fig8}?]}
			}
		\end{figure}

			In this Appendix, we show the projection of the band structure to the Ir $5d$ orbitals for type-II heterostructures in Fig.~\ref{fig8} and type-III of Mn and Co in Fig.~\ref{fig9}. The green shaded bands include high-energy four $j_{\rm{eff}}%$ 
			= 1/2$ bands and low-energy eight $j_{\rm{eff}}%$ 
			= 3/2$ bands. The results in Fig.~\ref{fig8} indicate that electrons are doped to the half-filled $j_{\rm eff}=1/2$ bands, realizing the spin-orbit coupled metallic states for all $A$ atoms. The doping rates are large (small) for $A$ = Mn and Co (Fe and Ni). Meanwhile, Fig.~\ref{fig9}(a) shows that the $j_{\rm eff}=1/2$ bands are slightly hole doped in the type-III with $A$ = Mn. Figure~\ref{fig9}(b) indicates that the type-III with $A$ = Co achieves an insulating state with $3/4$-filled $j_{\rm eff}=1/2$ bands. 
		
	\section{\label{appendix:C}Band structure of monolayer MgIrO$_3$}
\begin{figure}
		\includegraphics[width=0.9\columnwidth]{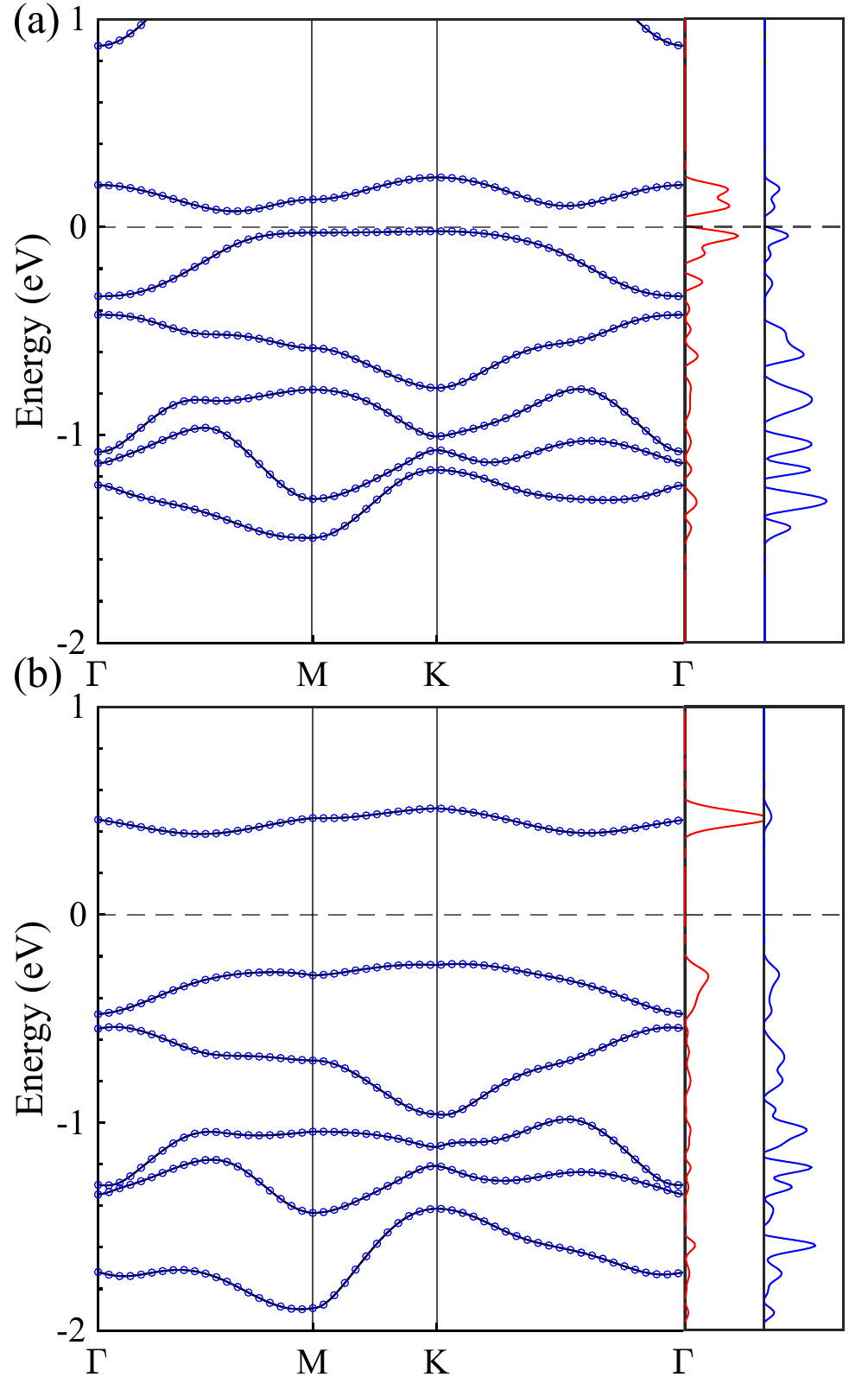}
		\caption{\label{fig10} Electronic band structure of monolayer MgIrO$_3$ obtained by the (a) LDA+SOC and (b) LDA+SOC+$U$ calculations for the in-plane N\'eel AFM state. In (b), the parameters of the Coulomb interaction and Hund's coupling are respectively set to $U_{\rm Ir}=3.0$~eV and $J_{\rm H}$/$U_{\rm Ir}=0.1$. The notions are common to Fig.~\ref{fig2}. %\red{[Please delete the values of $U$ and $J_{\rm H}/U$ above each figure, as in the other figures]}]
		}
	\end{figure}

		In this Appendix, we show the electronic band structures of monolayer MgIrO$_3$ obtained through {\it ab initio} calculations with the LDA+SOC [Fig.~\ref{fig10}(a)] and LDA+SOC+$U$ scheme [Fig.~\ref{fig10}(b)]. We set $U_{\rm Ir}=3.0$~eV and $J_{\rm H}/U_{\rm Ir} = 0.1$ in the LDA+SOC+$U$ calculations. In the LDA+SOC result, the system behaves as an insulating state with a tiny band gap of approximately $\sim$$0.096$~eV. %\red{[Is this an insulator?]} \blue{[Yes, it is an insulator.]} 
		However, the introduction of $U$ in the LDA+SOC+$U$ calculation results in a larger band gap, characteristic of the spin-orbit coupled insulator. We also calculate the PDOS of the $j_{\rm{eff}}=1/2$ and $3/2$ manifolds for Ir atoms, as shown in the right panels of Figs.~\ref{fig10}(a) and \ref{fig10}(b). The PDOS of Ir atoms certifies that the  $j_{\rm{eff}}=1/2$ and $j_{\rm{eff}}=3/2$ manifold are retained to support the spin-orbit coupled Mott insulating state in the monolayer, as in the bulk case~\cite{Jang2021}.
	
	% The \nocite command causes all entries in a bibliography to be printed out
	% whether or not they are actually referenced in the text. This is appropriate
	% for the sample file to show the different styles of references, but authors
	% most likely will not want to use it.

	\bibliography{my.bib}% Produces the bibliography via BibTeX.
	
\end{document}